\documentclass[prl,reprint,superscriptaddress,nobalancelastpage,twocolumn,showpacs]{revtex4-1}

\usepackage{color,amsthm,amsmath,amsfonts,graphicx,bm,amssymb}
\usepackage[plainpages=false]{hyperref}
\usepackage{epstopdf}
\usepackage[normalem]{ulem}
\usepackage{cancel}
\usepackage{comment}
\usepackage{etoolbox}
\usepackage{youngtab}
\usepackage{soul}
\usepackage{MnSymbol} 
\usepackage{mathtools}

\makeatletter
\newcommand{\specialcell}[1]{\ifmeasuring@#1\else\omit$\displaystyle#1$\ignorespaces\fi}
\makeatother

\newcommand{\beq}{\begin{equation}}
\newcommand{\eeq}{\end{equation}}

\def\genfun{J}

\def\kk{{\mathcal{K}}}
 
\def\ket#1{\mathinner{|{#1}\rangle}} 
\def\bbra#1{\mathinner{\llangle{#1}|}} 
\def\kket#1{\mathinner{|{#1}\rrangle}} 
\newcommand{\braket}[2]{\mathinner{\langle #1|#2\rangle}} 
\newcommand{\bbrakket}[2]{\mathinner{\llangle #1|#2 \rrangle}}

\def\Ai{\operatorname{Ai}}

\def\Proj{\Pi}
\def\Det{\operatorname{Det}}
\def\diagr{\xi}
\def\diagroth{\chi}
\def\tabl{{\bar\xi}}
\def\pv{\boldsymbol\mu}
\def\lv{\boldsymbol\lambda}
\def\xv{\mathbf{x}}
\def\yv{\mathbf{y}}
\def\av{\mathbf{a}}
\def\kv{\mathbf{k}}
\def\mv{\mathbf{m}}

\def\zero{\boldsymbol{0}}
\def\bv{\boldsymbol\beta}
\def\ZZ{\mathcal{Z}}
\newcommand{\pmatr}[2]{A_{#1}^{#2}}

\def\cbar{\bar c}
\def\partN{n}
\def\partM{m}
\def\partNS{n_s}
\def\QQ{\Lambda}

\def\Tr{\operatorname{Tr}}

\def\avemu{\hat{\mu}}
\def\sym{\operatorname{sym}}

\def\pp{p}
\def\dis{\eta}

\newcommand{\be}{\begin{equation}}
\newcommand{\ee}{\end{equation}}
\newcommand{\bea}{\begin{eqnarray}}
\newcommand{\eea}{\end{eqnarray}}

\newcommand{\nr}[1]{n_R(#1)} 
\newcommand{\nc}[1]{n_C(#1)} 
\def\Res{\operatorname{Res}}
\def\Sub{\operatorname{Sub}}

\newcommand{\titleinfo}{The crossing probability for directed polymers 
in random media.} 
\graphicspath{{../}, {../figures/}}
\begin{document}

\title{\titleinfo} 

\author{Andrea De Luca}
\email{andrea.deluca@lptms.u-psud.fr}
\affiliation{Laboratoire de Physique Th\' eorique et Mod\` eles Statistiques (UMR CNRS 8626), Universit\' e Paris-Sud, Orsay, France}

\author{Pierre Le Doussal}
\email{ledou@lpt.ens.fr}
\affiliation{Laboratoire de Physique Th\'eorique de l'ENS, CNRS \& Ecole Normale Sup\'erieure de Paris, Paris, France.}

\date{\today}

\begin{abstract}	
We study the probability that two directed polymers in the same random potential
do not intersect. We use
the replica method to map the problem onto the 
attractive Lieb-Liniger model with generalized statistics between particles.
Employing both the Nested Bethe Ansatz and known formula
from Mac Donald processes, we obtain analytical expressions for the first few moments
of this probability,
and compare them to a numerical simulation of a discrete model at high-temperature.
From these observations, several large time
properties of the non-crossing probabilities are conjectured.
Extensions of our formalism to more general observables are discussed.

\end{abstract}


\maketitle

\paragraph{Introduction. --- }

Recently there was considerable progress in calculating 
the free energy, and its fluctuations, for directed polymers, or
directed paths, in random media. This problem arises
in a variety of fields, including: optimization and glasses
~\cite{huse1985huse, *kardar1987scaling, *halpin1995kinetic},
vortex lines in superconductors~\cite{blatter1994vortices}, domain walls in magnets~\cite{lemerle1998domain}, disordered conductors \cite{SoOr07}, Burgers equation in fluid mechanics
\cite{bec}, 
exploration-exploitation tradeoff in population dynamics and 
economics \cite{dobrinevski} and in biophysics~\cite{hwa1996similarity,krug}.
Moreover, an exact mapping connects the DP 
in $1+d$ dimension to the 
Kardar-Parisi-Zhang (KPZ) equation~\cite{kardar1986dynamic}
in dimension $d$, 
which, in $d=1$, is at the center of an amazingly rich universality class, including 
discrete growth and particle transport models, with surprising
connections in mathematics to
random permutations and random 
matrices.

Two very different methods 
led to exact solutions: one based
on the limit of discrete lattices, e.g. particle models such as q-TASEP, often yielding
rigorous results \cite{png,spohnKPZEdge,corwinDP,borodin2014macdonald,Quastelflat};
 the other one based on replica, a standard approach in the
 physics of disordered systems \cite{kardareplica},
and the mapping 
to a 
continuum quantum integrable system, solvable by Bethe-ansatz
\cite{calabrese2010free,dotsenko,flat,SasamotoStationary}.
The calculation of the $\partN-$th moment of the DP
partition sum is reduced to the time-evolution
of a $n$-particle quantum state, determined by the initial conditions.
The evolution is performed with the 
attractive Lieb-Liniger Hamiltonian, whose spectrum is exactly computable \cite{ll,calabrese2007dynamics}. 
The derivation based on the replica-Bethe-ansatz (RBA) involves some guessing and has often anticipated
rigorous results from the math community. 
For instance, for the DP with two fixed endpoints, corresponding to the 
\textit{droplet initial condition} in the KPZ equation, both approaches obtain the free-energy 
as a Fredholm determinant, showing convergence at large time to the Tracy-Widom distribution 
for the largest eigenvalue of a random matrix \cite{calabrese2010free,dotsenko,spohnKPZEdge,corwinDP,calabreseSine}.

An outstanding challenge is to extend these methods and results to
collections of directed paths with hard-core repulsion, a 
difficult problem involving both interaction and disorder in
a non-perturbative way. It arises in the above examples,
e.g. populations competition, steps in vicinal surfaces
or the vortex glass in 2D superconductors \cite{natter}.
There was progress in that direction in the context of vortex arrays
\cite{EmigKardar}, 
within the multilayer PNG growth model \cite{ferrari}, and the
semi-discrete DP hierarchies \cite{borodin2014macdonald,Doumerc}, 
with emerging connections
to the spectrum of random matrices. Within the
RBA method, in almost all cases up to now, only the $1$d Bose-gas 
was considered, i.e. initial conditions corresponding to a fully symmetric quantum 
state. 
Here we consider infinite hard-core repulsion, modeled by a non-crossing condition,
which requires however more general initial conditions. 
\begin{figure}[t]
  \includegraphics[width=0.75\columnwidth]{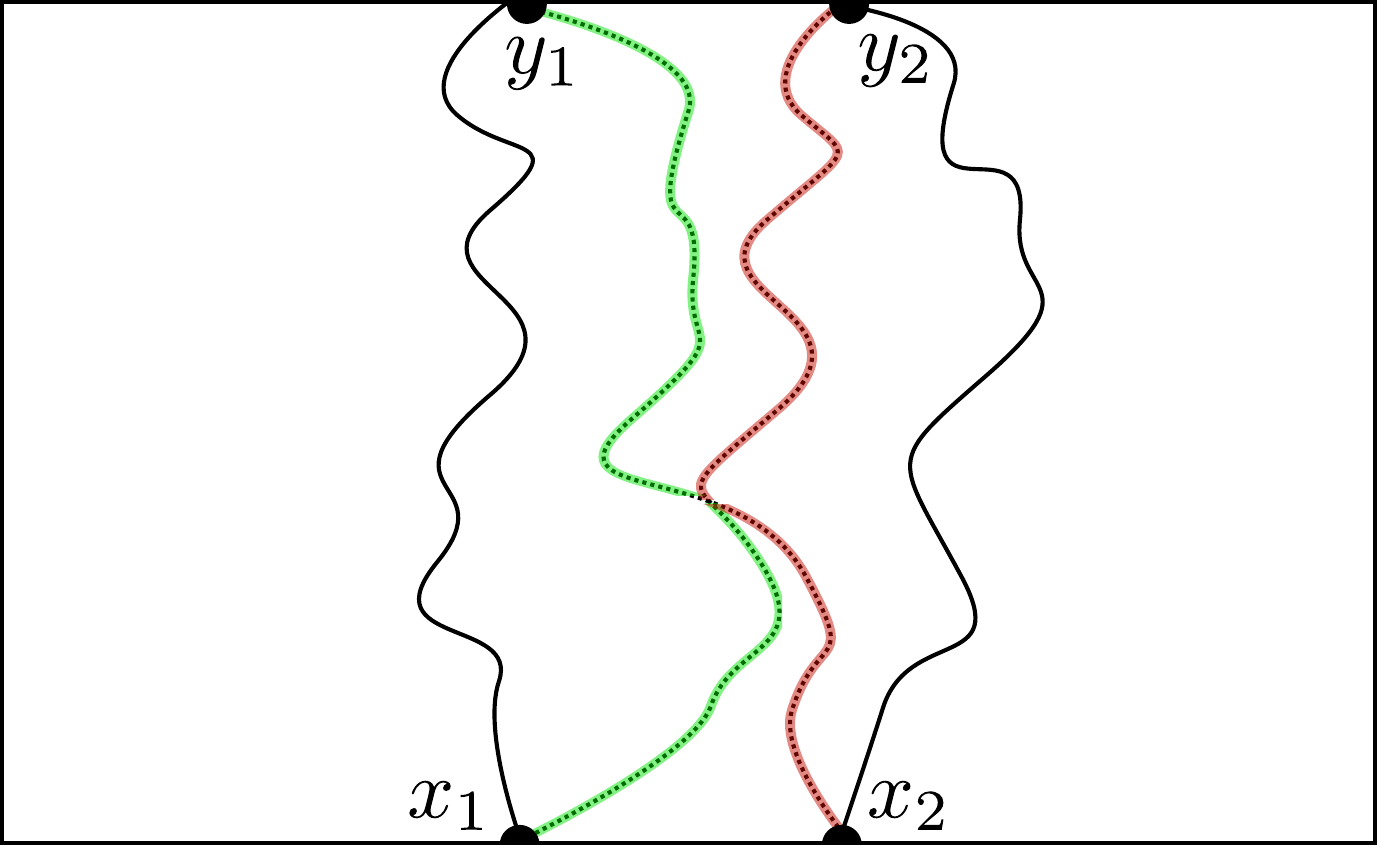}
  \caption{ 
  (Color online) Paths with fixed endpoints. By exchanging the paths after the last intersection, one builds
  a 
  mapping between crossing paths and paths with the 
  final endpoints exchanged.}
  \label{fig:Sketch}
\end{figure}

The aim of this Letter is to study continuum DP observables 
for non-crossing paths. We develop the more general nested
replica Bethe Ansatz (NRBA), and connect it to 
another recently developed method \cite{borodin2014macdonald}.
Here, as a first step, we focus on the calculation of
{\it crossing probabilities}, but we expect the 
potential outcome of the method to be broader.

We introduce the partition function of a directed polymer with fixed endpoints
\begin{equation}
 \label{Zorig}
 Z_\dis(x;y|t) \equiv \int_{x(0) = x}^{x(t) = y} Dx e^{-\int_0^t d\tau \left[ \frac{1}{4}(\frac{dx}{d\tau})^2 - \sqrt{2 \cbar} \dis(x(\tau), \tau)\right]} 
 \end{equation}
 in a random potential with white-noise correlations $\overline{\dis(x,t) \dis(x',t')} = \delta(x-x')\delta(t-t')$.  
  Then, the probability that two polymers with fixed endpoints do not cross in a given realization $\eta$
of the potential, is expressed as 
\begin{equation}
 \label{pnoncross}
 p_\dis(x_1,x_2;y_1,y_2|t) \equiv 1 - \frac{Z_\dis(x_2;y_1|t)Z_\dis(x_1;y_2|t)}{Z_\dis(x_1;y_1|t)Z_\dis(x_2;y_2|t)}
 \end{equation}
\noindent 
since all paths with at least one intersection can be obtained 
from paths 
with $y_1,y_2$ 
exchanged \cite{karlin1959coincidence, *gessel1985binomial, *gessel1989determinants, *wiki:lgvlem} (see Fig.~\ref{fig:Sketch}).
For simplicity, we will consider the random variable defined by the limit of near-coinciding endpoints
\begin{equation}
 \label{plimit}
 \pp_\dis(t) \equiv \lim_{\epsilon \to 0} \frac{p_\dis(-\epsilon, \epsilon; -\epsilon, \epsilon|t)}{4\epsilon^2}
 = \left.\partial_x \partial_y \ln Z_{\dis}(x; y|t)\right|_{\substack{x=0\\y=0}}
\end{equation}
where the last equality, derived from (\ref{pnoncross}), belongs to a larger set of relations between
non-crossing probabilities and the single path free energy \cite{Note1}
\cite{Note3}.
We now present a technique to calculate all 
the moments of $\pp_\dis(t)$ at arbitrary time $t$ 
with explicit results 
for the first few.
\paragraph{Replica trick and nested-Bethe ansatz. ---}
The average of products
$\ZZ_{\partN} = \overline{Z_\dis(x_1;y_1|t)\ldots Z_\dis(x_{\partN};y_{\partN}|t)}$ 
satisfies \cite{brunet2000probability}
\begin{equation}
\label{momentsZ}
\ZZ_{\partN}(\xv; \yv| t)  = \langle x_1 \ldots x_{n}| e^{-t H_{\partN}} |y_1 \ldots y_{n} \rangle 
\end{equation}
for any integer $\partN$, in quantum mechanical notations,
where bold symbols are shorthand for ordered sets of variables
and the Lieb-Liniger Hamiltonian 
reads:
\begin{equation}
\label{LLHam}
H_n \equiv -\sum_{i=1}^n \partial_{x_i}^2 + 2 c \sum_{1 \leq i<j \leq n} \delta(x_i - x_j) 
\end{equation}
with $c = -\cbar < 0$. 
To use the replica trick we introduce
\begin{equation}
 \label{plimitGen}
 \Theta_{\partN,\partM}(t) \equiv \lim_{\epsilon\to0} \overline{[(2\epsilon)^{-2} Z_\dis^{(2)}(\epsilon)]^\partM [Z_\dis(0;0|t)]^{\partN-2\partM}}
 \end{equation}
where we set $Z^{(2)}_\dis(\epsilon) \equiv Z_\dis(\epsilon;\epsilon|t)Z_\dis(-\epsilon;-\epsilon|t)-
Z_\dis(-\epsilon;\epsilon|t)Z_\dis(\epsilon;-\epsilon|t)$,
so that $\overline{\pp_\dis(t)^\partM} = \Theta_{0,\partM}(t)$. The advantage of this expression is that for integers 
$n,m$ with $\partN \geq 2\partM$,
it can be expressed in terms of \eqref{momentsZ}:
\begin{align}
 \label{pGenAve}
&\Theta_{\partN,\partM}(t) = \lim_{\epsilon\to0} (2\epsilon)^{-2\partM} \langle{\Psi_m}(\epsilon)|e^{-t H_{\partN}}|\Psi_m(\epsilon)\rangle = \nonumber \\
&= \sum_{\mu} \frac{|\mathcal{D}_m \psi_\mu(\zero)|^2}{||\mu||^2} e^{-t E_\mu}  
 \end{align}
where $|\Psi_m(\epsilon)\rangle = 2^{-\partM/2}(\otimes_{j=1}^\partM \ket{\epsilon,-\epsilon} - \ket{-\epsilon,\epsilon})\otimes\ket{0\ldots0}$ 
and a complete set of eigenstates $\ket{\mu}$ 
of $H_{\partN}$ of energies $E_\mu$ has been inserted with $\psi_\mu(\xv) \equiv \braket{\xv}{\mu} $. Here $\mathcal{D}_m$ is a differential operator
obtained from the limit $\epsilon\to0$ of $(2\epsilon)^{-m} \braket{\Psi_m(\epsilon)}{\mu}$, e.g. 
$\mathcal{D}_1 = \left.2^{-1/2}(\partial_{x_1} - \partial_{x_2})\right|_{\xv=0}$.  
Since $H_{\partN}$ is integrable by Bethe-ansatz,
the eigenstates, with eigenvalues $E_\mu=\sum_{j=1}^n \mu_j^2$, take the form 
\begin{equation}
 \label{BAwave}
  \psi_\mu(\xv) = \sum_{P,Q \in \mathcal{S}_{\partN}} \vartheta_Q(\xv) \pmatr{Q}{P} \exp\bigl[i \sum_{j=1}^n 
  x_{Q_j} \mu_{P_j}\bigr] 
\end{equation}
where $\{\mu_1,\ldots, \mu_n\}$ is a set of rapidities,
$\mathcal{S}_{\partN}$ 
the set of $\partN$-permutations and $\vartheta_Q(\xv)$ 
the indicator of the sector $x_{Q_1} \leq x_{Q_2} \ldots \leq x_{Q_{\partN}}$. 
However, $|\Psi_m(\epsilon)\rangle$ is not a symmetric state under the exchange 
of the coordinates, thus the quantum dynamics described by \eqref{LLHam} does not belong to the bosonic sector. 
Nonetheless, it is still possible to explicitly determine the eigenstates \cite{NBAref}, 
corresponding to different representations of the symmetric group. It is enough to choose
the vectors $\pmatr{Q}{P}$, for all fixed permutation $P$, inside an irreducible representation of 
$S_{\partN}$. The relevant case for us is the representation corresponding to a two-rows Young diagram $\diagr = (\partN-\partM, \partM)$, 
where we 
denote a diagram as the decreasing sequence of row lengths \cite{fulton1991representation}.
For instance, for $\partN=8$ and $\partM=3$ we have
\begin{equation}
\label{youngtabex}
\Yvcentermath1
 (5,3) \equiv \young(13579,246) \;. 
\end{equation}
and the filling indicates antisymmetric wave-functions under the exchange of coordinates $\mathinner{x_1 \leftrightarrow x_2},\mathinner{x_3 \leftrightarrow x_4},\mathinner{x_5 \leftrightarrow x_6}$,
which are in the 
symmetry class selected by the action of $\mathcal{D}_{m=3}$.
These 
representations can be built explicitly as the Hilbert space
of an integrable spin-$1/2$ chain with $\partN$ sites restricted to the sector
with $\partM$ down spins. 
Then the eigenstates of $H_n$ on a ring of length $L$ are obtained 
diagonalizing simultaneously the spin-model. This leads to
the so-called nested-Bethe-ansatz (NBA) equations 
\begin{subequations}
\label{NBAeq}
\begin{align}
&\prod_{\substack{b=1\\b\neq a}}^\partM \frac{\lambda_{ab} - i c}{\lambda_{ab} + i c} = 
\prod_{j=1}^\partN \frac{\lambda_a - \mu_j - i c/2}{\lambda_a - \mu_j + i c/2}\;, \label{NBAeqla}
\\
&\prod_{\substack{k=1\\k \neq j}}^{n} \frac{\mu_{jk} + i c}{\mu_{jk} - ic}
\times \prod_{a=1}^{m}
\frac{\mu_j - \lambda_a - i c/2}{\mu_j  - \lambda_a + i c/2}
= e^{i \mu_j L} \label{NBAeqmu}
\end{align}
\end{subequations}
where $\mu_{\alpha\beta} = \mu_{\alpha} - \mu_{\beta}$ and same for the $\lambda$'s, the auxiliary rapidities on the spin chain that impose the appropriate symmetry to the wave-function. Solutions
of (\ref{NBAeq}) provide the eigenstates of \eqref{LLHam} in the appropriate symmetry class and the wave-functions are obtained setting
$\pmatr{Q}{P} = \pmatr{\text{\tiny bos}}{P}\bbrakket{\Psi_Q}{\omega(P\pv)}$ with 
\begin{equation}
 \label{APQSpinChain}
\kket{\omega(\pv)} = \sum_{a_1,\ldots,a_m = 1}^\partN \alpha(\av| \pv ) 
\sigma_{-}^{a_1}\ldots \sigma_{-}^{a_m}\kket{+} \;.
\end{equation}
Here, $(P \mu)_j \equiv \mu_{P_j}$ and $\kket{\ldots}$ indicates states in the auxiliary spin space, $\sigma_{-}^a$ 
is the lowering spin operator at site $a$, $a=1,\ldots,m$,
acting on the reference state $\kket{+} = \kket{\uparrow\ldots\uparrow}$. 
The vector of states $\kket{\Psi_Q}$ is fixed by the filling of $\diagr$ and 
performs the unitary mapping between the spin-chain representation and 
a particular representation of shape $(\partN-\partM,\partM)$, such that
the exchange of two-spins is mapped into the exchange of two particles.
Here $\pmatr{\text{\tiny bos}}{P} = \Omega_{\pv}^{0}/\Omega_{P\pv}$ 
accounts for the bosonic phase scattering with 
$\Omega_{\pv} \equiv \prod_{\substack{j<l}} f(\mu_{lj})$,
$\Omega_{\pv}^{0}=\sqrt{\Omega_{\pv}  \Omega_{-\pv}}$ 
and $f(u) \equiv u/(u-i c)$, while 
\begin{align*} 
 \alpha(\av| \pv ) &= \sym_{\lv}\left[\prod_{k<l}\left(1 + \frac{i c \operatorname{sgn}(a_{lk})}{\lambda_{lk} }\right) 
  \prod_{l=1}^\partM \kappa_{a_l} ( \lambda_{l}| \pv )\right] \;, \\
  \kappa_a(u | \pv  ) &= \frac{i c}{u - \mu_a  - i c/2} \prod_{b=a}^\partN \frac{u - \mu_b - i c/2}{u - \mu_b + i c/2}\;.
\end{align*}
and $\sym_{\lv}[W(\lv)] = \sum_{R} W(R\lv)/\partM!$ is the symmetrization of $W(\lv)$ 
over the variables $\lv$.

\paragraph{Average of $p_\dis(t)$. ---} We now consider the case $\partM=1$ which selects 
the subspace of wave-functions $\Psi_{\pv}(x_1 = x_2) = 0$. 
Then, the wave-function in \eqref{BAwave}
remains continuous, even after the action of $\mathcal{D}_1$, and we 
can average 
over different orderings of the coordinates $\xv$.
Hence $\mathcal{D}_1 \psi_\mu(\zero) = \frac{1}{n!} \sum_{P,Q} d_1(Q^{-1} P \pv) \pmatr{Q}{P}$, where 
$d_\partM(\pv) = (-i)^\partM \mathcal{D}_\partM e^{i \pv \cdot \xv}|_{\xv = 0}$. We then obtain 
\cite{Note4}
\begin{equation}
\label{sumQ}
\frac{1}{n!} \sum_Q d_1(Q^{-1} \pv) \kket{\Psi_Q} = \sqrt{Z} {\sum_{a=1}^n} (\mu_{a} - \avemu) \sigma_-^a \kket{+} 
\end{equation}
with 
$Z = [(\partN-1)\partN!]^{-1}$
ensuring normalization and $\avemu = \sum_b \mu_b /\partN$. It leads to 
$ | \mathcal{D}_1 \psi_\mu(\zero) |^2 = \partN(\partN-2)!|\Omega_{\pv}|^2  |\mathcal{F}(\lambda, \pv)|^2 $, 
where 
$\mathcal{F}(\lambda, \pv) = \sym_{\pv}\left[\Omega_{\pv}^{-1} \sum_a (\mu_{a} - \avemu) \kappa_a(\lambda | \pv  )\right]$ and we note $\lambda_{a=1}=\lambda$.
The sum over all solutions of \eqref{NBAeq} must then be performed according to \eqref{pGenAve}, a formidable task in general. 
However, \eqref{NBAeqmu} simplifies dramatically when $L\to\infty$. For $\cbar>0$,
the $n$ rapidities $\mu_1,\ldots, \mu_n$,  are organized in $\partNS$ bound states,
each composed by $m_j\geq 1$ particles, with $\sum_{j=1}^{\partNS} m_j = n$. 
The rapidities inside a bound state follow a regular pattern in the complex plane 
$\mu^{j,a} = k_j + \frac{i \cbar}{2} (m_j + 1 - 2a) + i \delta^{j,a}$, named string. 
Here $a=1,\ldots, m_j$ labels the rapidities inside the string and 
$\delta^{j,a}$ are exponentially small 
for large $L$. 
A study of \eqref{NBAeq} reveals that, at variance with the bosonic case, 
not all string configurations are actually allowed, 
consistently with the symmetry of the wave-function \cite{Note1}.
For those allowed, using \cite{calabrese2007dynamics,pozsgay2012form}, we 
obtain their norm \cite{Note5} as:
\begin{align}
 \frac{||\psi_\mu||^2}{(\Omega_{\pv}^0)^2} &= 
 \frac{ (L \cbar)^{\partNS} \prod_{j} m_j^2 }{\cbar^{\partN}\Phi(\kv, \mv)}{\sum_{l=1}^n} \frac{4c^2}{c^2 + 4(\mu_l - \lambda)^2}\;,\\
 \Phi(\kv, \mv) &= \prod_{1\leq j<j'\leq \partNS} \frac{(k_j - k_{j'})^2 + c^2(m_j - m_{j'})^2/4}{(k_j - k_{j'})^2 + c^2(m_j + m_{j'})^2/4}\;.
 \end{align}
For each configuration of rapidities following the string ansatz, a multiplet of eigenstates is given by the set $\{\lambda^{(1)},\ldots,\lambda^{(\partN-1)}\}$ 
of solutions of \eqref{NBAeqla}, 
i.e. $Q(\lambda) \equiv P_{+}(\lambda)/P_{-}(\lambda) = 1$, where $P_{\pm}(\lambda) = \prod_i (\lambda-\mu_i\pm ic/2)$. 
These values cannot be determined
analytically for general $n$, however the sum over them can be performed using residue theorem
$$ \sum_{i} w(\lambda^{(i)}) = \oint_\mathcal{C} \frac{dz}{2\pi i} w(z) \frac{Q'(z)}{Q(z)-1} $$
where $w(z)$ is any analytic function inside the contour $\mathcal{C}$, which encircles all the solutions $\lambda^{(i)}$ 
and no other singularity of the integrand. Equivalently, the integral can be computed
taking the poles outside the contour, which in the case $w(z) = \frac{|\mathcal{F}(\lambda^{(i)},\pv)|^2}{||\psi_\mu||^2}$, 
are given by $z_k = \mu_k - ic/2$ \cite{Note5}. 
The sum can then be performed 
analytically. Moreover, for $L\to \infty$, string momenta become free and we can replace $\sum_{k_j} \to m_j L \int \frac{d k_j}{2\pi}$, which leads to 
\begin{align}
\label{QpExp}
& \Theta_{\partN,1}(t) =  \sum_{\partNS=1}^\partN \frac{\partN! \cbar^\partN}{\partNS! (2 \pi \cbar)^{\partNS}} \sum_{(m_1,..m_{\partNS})_{\partN}} \\ 
& \prod_{j=1}^{\partNS} \int_{-\infty}^{+\infty} \frac{dk_j e^{-A_2 t}} {m_j} 
\Phi(\kv, \mv) \QQ_{\partN,1}(\kv, \mv)  \nonumber
\end{align}
with $ \QQ_{\partN,1} = (\partN(\partN-1))^{-1} h_2$. 
Here, $(m_1, \ldots ,m_{n_s})_n$, indicates sum over all 
integers $m \geq 1$ whose sum equals $\partN$ and 
we defined $ h_p = \sum_{j<l} (\mu_j-\mu_l)^p - (i\cbar)^p  (j-l)^p$.
The rapidities $\mu_j$ are written as a function of string sizes and momenta according to the string ansatz, 
so that $\QQ_{\partN,1}$ vanishes on the $n$-strings.
The conserved charges of the Lieb-Liniger model have been introduced as $A_p = \sum_{j=1}^\partN \mu_j^p$, $A_2$ being the energy. 
A crucial property of \eqref{QpExp} is that by replacing
$\QQ_{\partN,1} \to \QQ_{\partN,0} \equiv 1 $, one recovers the formula for 
$\ZZ_{\partN}(t)\equiv \ZZ_{\partN}(\xv = \zero;\zero| t) = \Theta_{n,0}(t)$ as given in \cite{calabrese2010free}. Therefore, rewriting $\QQ_{\partN,1}$
in terms of the conserved charges and using the statistical tilt symmetry (STS) (see e.g. Appendix of \cite{flat}), we obtain
\begin{equation}
\Theta_{\partN,1}(t) = \frac{1}{\partN-1} \left[\frac{\partN (\partN^2-1) \cbar^2}{12}  - \partial_t - \frac{1}{2t} \right] \ZZ_{\partN}(t) \label{eq1}  \;.
\end{equation}
This expression is exact for $\partN \geq 2$ and allows the analytical continuation $\partN \to 0$. In particular, we obtain 
\begin{equation}
\label{pfinal0}
 \overline{p_\dis(t)} = \lim_{\partN \to 0} \Theta_{\partN,1}(t) = \frac{1}{2t} \;.
\end{equation}
This is in fact the exact result for $p_\dis (t)$ without disorder, i.e. $\eta(x,t) = 0$.
This remarkable conclusion can also be obtained 
by averaging \eqref{plimit} and recalling that the dependence of 
the {\it average} free energy of a path with respect to its endpoints is 
entirely fixed by the STS, namely 
$\overline{\ln Z_\eta(x; y|t)} = h(t) - (x-y)^2/(4 t)$, where
\begin{equation} \label{fe}
h(t) = \overline{\ln Z_\eta(0; 0|t)}
\end{equation}
is our averaged free energy (and average KPZ height).

\paragraph{Alternative derivation. ---}
A different approach was recently proposed in \cite{borodin2014macdonald}
(remark 5.25) 
where non-intersecting paths were also studied. There, it was proposed
a multicontour-integral
formula associated to a partition of $\partN$. We identify the partition with a Young-diagram and 
for the two-row case of our interest, it can be put in the form
\begin{multline}
 \label{BorodinContour}
 \Theta_{\partN,\partM}(t)= \frac{1}{2^\partM}\int \frac{dz_1}{2\pi}\cdots \int\frac{dz_{\partN}}{2\pi} 
  e^{- t  \sum_{k=1}^\partN z_{k}^2}
   \\ \times  
\left(\prod_{1 \leq  k < j\leq \partN} f\bigl(z_{kj}\bigr)\right)
 \left(\prod_{q=1}^\partM h(z_{2q-1,2q})
 \right)
 \;.
 \end{multline}
where $z_{kj}= z_k-z_j$, $h(u) = u(u-ic)$ and 
the integration contours are parallel to the real axis with an imaginary part $C_j$ for $z_j$ 
satisfying $C_{j+1} > C_{j}+\cbar $.
Shifting back all the contours to the real axis,
we encounter many poles
whose residues reduce to integrals with a smaller number of integration variables.
This expansion can then be organized 
to reproduce the one based on strings in \eqref{QpExp}, with $\QQ_{\partN,1}$ replaced by \cite{Note1}
 \begin{equation}
 \label{borodinG}
  \QQ_{\partN,\partM}(\kv,\mv) = \frac{1}{2^\partM}\sym_{\pv} \left[\frac{\prod_{q=1}^\partM h(\mu_{2q-1,2q})}{\prod_{1\leq j<k\leq \partN} f(\mu_{kj})}
  \right]
 \end{equation}
and again the $\pv$ given by the string ansatz. 
Interestingly, $\QQ_{\partN,\partM}$ is always a polynomial in the $\mu$'s of degree $2m $
as can be seen considering the residue at coinciding points.
Moreover, Eq.~\eqref{borodinG} agrees with the result obtained from the NBA for $m=1$,
which gives a completely independent check to the proposition 
in \eqref{BorodinContour}. 
For $m > 1$, the calculation from NBA becomes more involved but we will continue by assuming that \eqref{borodinG} retains its validity. 
\begin{figure*}[ht]
\begin{minipage}{0.495\textwidth}
\centering
\includegraphics[width=0.9\textwidth]{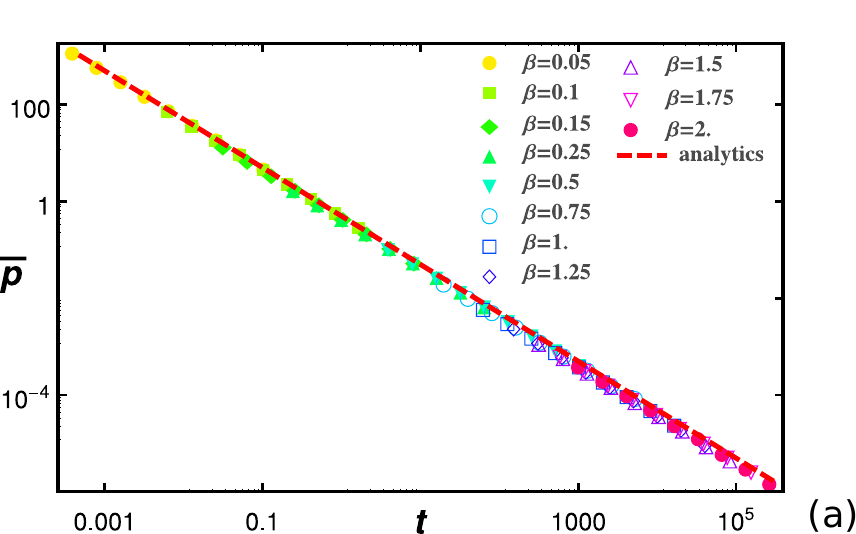}
\end{minipage}
\begin{minipage}{0.495\textwidth}
\centering
\includegraphics[width=0.9\textwidth]{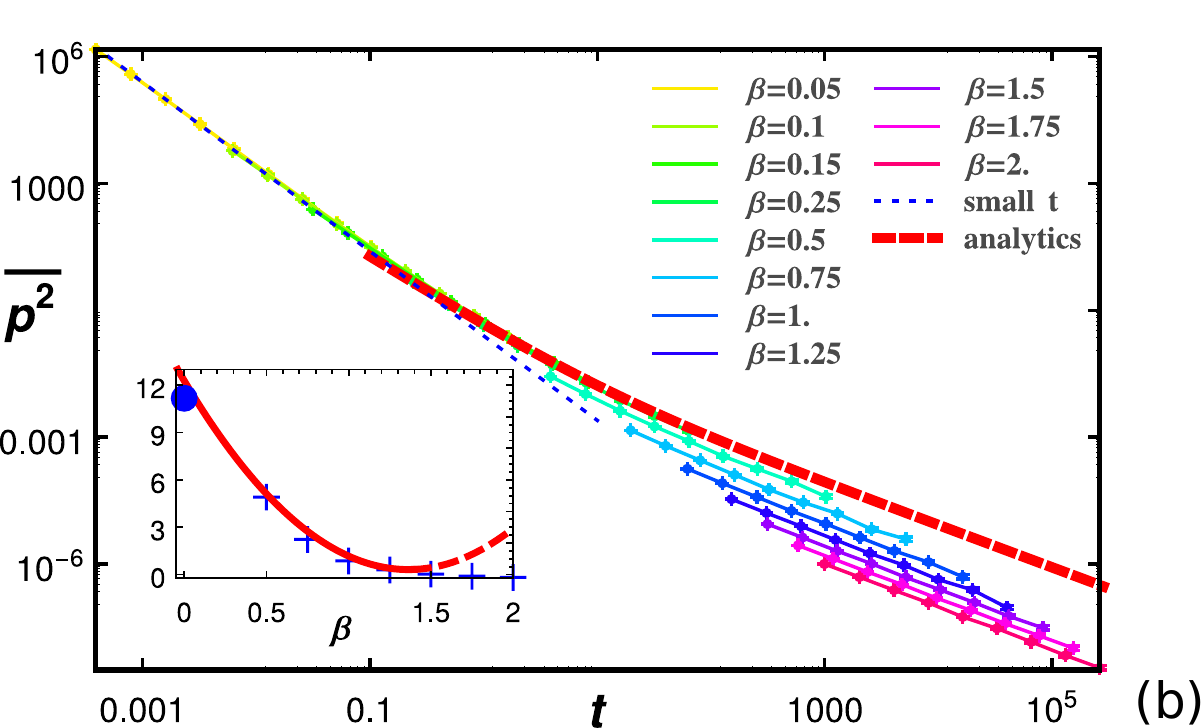}
\end{minipage}
  \caption{(Color online) The first \textbf{(a)} and second moment \textbf{(b)} of $p_\eta$ in the continuum limit
  are shown vs $t$ for several value of $\beta$. 
  Numerical simulations with at least $2\times 10^5$ realizations and time 
  up to $\hat t = 8192$.
  The value of $p_\eta$ introduced in \eqref{plimit} is obtained by $p_\eta(t) = \hat{p}/(16\beta^4)$, with $\hat{p}$ the
  probability on the lattice and the time is scaled to $t$ by $\hat t = t/(2\beta^4)$. 
  Dashed red line: analytical predictions, \eqref{pfinal0} and \eqref{p2final0}.
  In \textbf{(b)}, the small $t$ expansion $\overline{p_\eta^2}\simeq 4t^{-2}$ 
  is shown with a dashed blue line. Inset in \textbf{(b)}: plot of 
  $\overline{\hat p^2}\, \hat t/\beta^4 \simeq \beta^{-4} c_2(\beta)$
  vs $\beta$ and extrapolation at small $\beta=0$ from
  the best fit with a quadratic function in $\beta$
  at fixed $t=10^3$. It shows a finite limit consistent with our prediction $\simeq 32/3 = 10.6(7)$ (blue circle).
  }
  \label{fig:numerics}
\end{figure*}
\paragraph{Higher moments of $p_\dis(t)$. ---} 
We now focus on $\partM=2$. 
Upon symmetrization in \eqref{borodinG} one obtains
\begin{equation}
  \QQ_{\partN,2} (\pv) = 
\frac{h_2^2 - (\partN-1)h_4 - \partN(\partN-1)^2 \cbar^2 h_2 }{\partN(\partN-1)(\partN-2)(\partN-3)}\;.
\end{equation}
After 
tedious calculations, 
it can 
be rewritten in terms of the conserved charges $\QQ_{\partN,2}(\{A_p\})$ \cite{Note1}. 
In contrast with the $m=1$ case, higher charges, up to $p=4$, are involved. It is therefore useful to formally generalize the partition function to
$\ZZ_\partN^{\bv}(t)$ which is obtained from the expression of $\ZZ_\partN(t)$ replacing the imaginary time evolution $e^{-A_2 t}$ 
with the more general $e^{-A_2 t + \sum_{p \geq 1} \beta_p A_p}$.
This is the partition function of a Generalized Gibbs ensemble (GGE) \cite{fioretto2010quantum},
which we show can be related to
a Fredholm determinant \cite{Note1}. Here, we use it as a generating function: 
$\Theta_{\partN,2}(t) = \QQ_{\partN,2}(\{\partial_p\})\ZZ_\partN^{\bv}(t)$, 
formally replacing $A_p \to \partial_p \equiv \partial_{\beta_p}$ and setting $\beta_p \to 0$ at the end. Deriving extended
STS identities from the invariance $\mu_j \to \mu_j + k$ in $\ZZ_\partN^{\bv}(t)$, for arbitrary $k$,
we are able to re-express it only from the energy $A_2$, leading to
%
\begin{equation}
\label{p2eq}
\overline{p_\dis(t)^2} = 
- \left(\frac{1}{t} \partial_t + \frac{1}{2} \partial_t^2 \right) h(t) \;.
\end{equation}
Hence the second moment is determined at all times from 
the average free energy $h(t)$ (\ref{fe}). We did not
find a direct derivation of this remarkable result, and
it may be a consequence of integrability. 
At large time $h(t) \simeq -\frac{\cbar^2 t}{12} + \overline{\chi_2}(\cbar^2 t)^{1/3}$
 \cite{calabrese2010free,dotsenko,spohnKPZEdge,corwinDP}, so that
\begin{equation}
 \label{p2final0}
 \overline{p_\eta(t)^2}  \simeq \frac{\cbar^2}{12 t} -  \frac{2 \overline{\chi_2} \cbar^{2/3}}{9 t^{5/3}}
\end{equation}
with $\overline{\chi_2}=-1.77(1)$ the mean of the Tracy-Widom GUE distribution \cite{tracy1994level}. 
Repeating this procedure for $m=3$ we can again use $\ZZ_{\partN}^{\bv}(t)$. Now
higher charges are involved and the result is expressed as derivatives of
a Fredholm determinant \cite{Note1}. It simplifies at large time leading to:
\begin{equation}
 \label{p3final0}
 \overline{p_\eta(t)^3}  \simeq \frac{\cbar^4}{15 t} -  \frac{2 \overline{\chi_2} \cbar^{8/3}}{9 t^{5/3}}
\end{equation}
It is natural to conjecture the leading decay $\overline{p_\eta(t)^m}  \simeq \gamma_m \cbar^{2(m-1)}/t$
for any integer $m>1$. 
However, the knowledge of moments at long-times is not sufficient to
reconstruct the full distribution of $p$:
in view of \cite{Doumerc}, 
we further surmise that $p_\eta(t)$ tends to zero 
(sub) exponentially at large $t$ for all but a small fraction $\sim 1/(\bar c^2 t)$ of 
environments where typically $p_\eta(t) \sim \cbar^2$. 
This is consistent with the conjecture \cite{Note1} 
$\overline{\ln p_\eta(t)} \sim - a  (\bar c^2 t)^{1/3}$ where
$a=\overline{\chi_2} - \overline{\chi'_2}$ is the average gap between the first ($\chi_2$)
and second ($\chi'_2$) GUE (scaled) largest eigenvalues, with
$a \approx 1.9043$ \cite{TWNote} (note that $a \approx 1.49134$ for
the hard wall problem \cite{Gueudre}).

\paragraph{Comparison with numerics ---}
To check our results, we study a discrete directed polymer on a square lattice \cite{calabrese2010free},
defined according to the recursion (with integer time $\hat t$ running along the diagonal) 
\begin{equation}
 Z_{\hat x, \hat t + 1} =  (Z_{\hat x - \frac{1}{2},t} + Z_{\hat x + \frac{1}{2}, \hat t} ) e^{-\beta V_{\hat x, \hat t + 1}}
\end{equation}
with $V_{\hat x, \hat t}$ sampled from the standard 
normal distribution. In the high temperature limit $\beta \ll 1$,
it maps into the continuous DP \eqref{Zorig} at $\cbar = 1$ with $x = 4\hat x \beta^2$ and $t = 2\hat t \beta^4$ 
\cite{calabrese2010free}. We consider 
two polymers with initial conditions $Z_{\hat x, \hat t = 1}^{\pm} = \delta_{\hat x, \pm 1/2}$
and ending at time $\hat t$ at 
$\hat x = \pm 1/2$.
For each realization of the 
$V_{\hat x, \hat t}$, the non-crossing probability $\hat p$ on the lattice 
is efficiently computed using the image method 
\cite{karlin1959coincidence, *gessel1985binomial, *gessel1989determinants, *wiki:lgvlem}.
Comparing with \eqref{plimit}, we deduce $\hat p \simeq 16 p \beta^4$, for $\beta \to 0$, due to the rescaling of the factor $\epsilon = 4\beta^2$.
The numerical results 
and the analytical predictions (\ref{pfinal0}, \ref{p2final0}) are shown 
in Fig.~\ref{fig:numerics}. For the first moment $\overline{p_\eta(t)}$ the agreement is
excellent even at finite temperature, presumably due to robustness of 
(\ref{pfinal0}) at any time. 
The numerical check of (\ref{p2final0})
is more delicate:
indeed, the large-time behavior of the second 
moment depends strongly on temperature and approaches our prediction 
only for $\beta \ll 1$, see Fig.~\ref{fig:numerics} (right).
However, the leading decay is found consistent with $t^{-1}$ 
down to zero temperature, where the 
polymer paths do not fluctuate thermally
and for any $\partM>0$: $\overline{\hat p^\partM}=\overline{\hat p}$.
In order to interpolate between zero and high temperatures, we conjecture the large
time behavior of the moments on the lattice: $\overline{\hat p^\partM} \simeq  c_\partM(\beta)/\hat{t}$, 
with $c_\partM(\beta) \simeq \gamma_m 2^{4m-1} \beta^{4(\partM-1)}$ at high temperatures and $c_\partM(\beta\to\infty)=c_\infty$
is a constant 
that we expect 
to be non-universal \cite{Note1}. 
This agrees with the intuitive picture of the zero temperature deterministic path, weakly perturbed by thermal fluctuations. Checking the sub-leading terms in (\ref{p2final0})
would require much more intensive numerics. 

\paragraph{Conclusions. ---}
We presented a general formalism to
calculate the statistics of $N$ mutually avoiding directed polymers in a random potential,
with explicit results for $N=2$. Multi-polymer observables are reduced to a compact form in terms 
of conserved quantities of the Lieb-Liniger Hamiltonian and expressed at all times
by derivatives of a Fredholm determinant, i.e. the GGE partition function.
As a simplest example we obtained the lowest moments of non-crossing 
probabilities, with an exact relation between the variance and the free-energy,
a non trivial scaling $t^{-5/3}$ for the sub-leading part and a prediction of
leading behavior for all moments. The full distribution of 
the non-crossing probability is under current investigation. 
Going beyond the infinite hard-core repulsion remains for the moment elusive,
but 
we are confident that 
further developments of the present method
and full exploitation of its integrable structure,
will allow further progress in the elusive interplay
between disorder and interactions. 


\paragraph{Acknowledgements. ---}
We thank A. Borodin, I. Corwin and A. Rosso for fruitful discussions. 
This work is supported by ``Investissements d'Avenir'' 
LabEx PALM (ANR-10-LABX-0039-PALM)
and by PSL grant ANR-10-IDEX-0001-02-PSL.

\input{main.bbl}

\clearpage
\newpage

\clearpage 
\setcounter{equation}{0}%
\setcounter{figure}{0}%
\setcounter{table}{0}%
\renewcommand{\thetable}{S\arabic{table}}
\renewcommand{\theequation}{S\arabic{equation}}
\renewcommand{\thefigure}{S\arabic{figure}}

\onecolumngrid

\begin{center}
{\Large Supplementary Material for EPAPS \\ 
\titleinfo
}
\end{center}
Here we give additional details about 
\begin{itemize}
\item crossing probabilities of paths;
 \item the construction of the representations of the symmetric group;
 \item the explicit derivation of the $\partM=1$ from the nested-Bethe ansatz;
 \item the generalization to arbitrary initial conditions from the expansion of the multi-contour integral formula;
 \item the derivation of a Fredholm determinant formula for the generalized generating function.
 \item the extension of the STS symmetry and the calculation of the $\partM=1,2,3$ moments;
\end{itemize}
We will conventionally use  $\diagr$ (and $\diagroth$) to indicate Young diagrams of size $\partN$, 
with $\xi_i$ ($\xi^i$) the size of rows (columns) from top to bottom (left to right). 
We write each diagram as the sequence of decreasing row sizes $\xi \equiv (\xi_1 \xi_2\ldots)$,
 $\xi_1 \geq \xi_2.. \geq \xi_{n_R(\xi)}$. Equivalently, one can write it as a 
sequence of decreasing column sizes. There is a correspondence (bijective in each case) 
between these two sets of sequences, and the 
partitions of the integer $\partN$, as $\sum_{\alpha=1}^{\nr{\xi}} \xi_\alpha = \sum_{\alpha=1}^{\nc{\xi}} \xi^\alpha 
=\partN$, with 
$\xi_\alpha \geq \xi_{\alpha'}$ when $\alpha<\alpha'$ and the same for $\xi^\alpha,\xi^{\alpha'}$.
A standard tableau $\tabl$ is obtained from the diagram $\diagr$ by filling the boxes with 
the integers $1,\ldots, \partN$, with the condition that they grow from left/right and top/bottom along each row and column.

\section{Crossing probabilities and the free energy}

Here we show relations, valid in any given disorder realization, 
between non-crossing probabilities of several paths
and the free energy of a single path. Consider $N$ directed paths $x_i(\tau)$ in the continuum, with endpoints $x_i(0)=y_i$, $x_i(t)=y_i$
and $x_1<x_2<..x_N$, $y_1<y_2<..y_N$. We know from the Karlin-McGregor formula and generalizations 
\cite{karlin1959coincidence, *gessel1985binomial, *gessel1989determinants, *wiki:lgvlem}
that the probability that these paths do not cross can be expressed as a determinant:
\bea
p_\eta^{(N)}(x_1,\ldots, x_N;y_1,\ldots, y_N| t) = \prod_{i=1}^N \frac{1}{Z_\eta(x_i;y_i|t) } {\rm det} [ Z_\eta(x_i;y_j|t) ]_{N \times N} 
\eea 
Consider first $N=2$ which is the main focus of this work. Bringing two endpoints close together, i.e. 
$y_{1,2} = y \mp \epsilon/2$ and dividing by $\epsilon$, equivalently applying
the operator $O_2(y)=\frac{1}{2} (\partial_{y_2}-\partial_{y_1})|_{y_2=y_1=y}$, or
equivalently for $x_{12}$ we obtain:
\bea
&& p^{(2)}_\eta(x_1,x_2;y|t) = \partial_y ( \ln Z_\eta(x_2;y|t) - \ln Z_\eta(x_1;y|t) ) \\
&& p^{(2)}_\eta(x;y_1,y_2|t) = \partial_x ( \ln Z_\eta(x;y_2|t) - \ln Z_\eta(x;y_1|t) ) 
\eea 
Bringing points together on both ends we obtain the 
non-crossing probability for 2 paths both from $y$ to $x$:
\bea
p^{(2)}_\eta(x;y|t)  := O_2(x) O_2(y) p^{(2)}_\eta(x_1,x_2;y_1,y_2|t)
= \partial_x \partial_y  \ln Z_V(x;y|t) 
\eea 
In particular this leads to Eq. (\ref{plimit}) for the quantity defined in the text 
$p_\eta(t) := p^{(2)}_\eta(0;0|t)$. Note that the STS symmetry also implies
\bea
&& \overline{p^{(2)}_\eta(x_1,x_2;y|t)} = \frac{x_2-x_1}{2 t} \quad , \quad p^{(2)}_\eta(x;y_1,y_2|t) 
= \frac{y_2-y_1}{2 t} \quad , \quad \overline{p^{(2)}_\eta(x;y|t)} = \frac{1}{2 t}
\eea 
in addition to the result mentioned in the text. Note that these STS results are
exact also for a model with a more general noise, provided it has the STS symmetry,
e.g. $\overline{\eta(x,t) \eta(x',t')} = \delta(t-t') R(x-x')$. 

Similar relations, although more involved, exist for more than 2 paths, $N>2$. 
For instance the non-crossing probabilities with one endpoint coinciding,
defined as
$p_\eta^{(N)}(x;y_1,..y_N,t) = O_N(x) p_\eta^{(N)}(x_1,..x_N;y_1,..y_N,t)$
where now $O(x) = \frac{1}{N!} \prod_{1 \leq i<j \leq N} (\partial_j-\partial_i)|_{x_a=x}$,
admits a simple expression as a determinant:
\bea
&& p_\eta^{(N)}(x;y_1,..y_N | t) = det [  \frac{\partial_x^{i-1} Z_\eta(x,y_j)}{Z_\eta(x,y_j)} ]_{N \times N} 
\eea 
the first row of the matrix being the vector $(1,..,1)$. Such derivatives can then be re-expressed
in terms of derivatives of the free energy. Taking the second endpoint coinciding by applying $O_N(y)$ leads to,
e.g. for $N=3$:
\bea
&& p^{(3)}_\eta(x;y|t) = p^{(2)}_\eta(x;y|t) \partial_x \partial_y p^{(2)}_\eta(x;y|t) 
- \partial_x p^{(2)}_\eta(x;y|t) \partial_y p^{(2)}_\eta(x;y|t) + 2 p^{(2)}_\eta(x;y|t)^3 
\eea 
where we recall $p^{(2)}_\eta(x;y|t) = \partial_x \partial_y \ln Z_\eta(x;y|t)$, and to more complicated relations for higher $N$. \\

{\it Dependence in elastic coefficient and a conjecture for discrete model:}

If, in the continuum model described by Eq. (\ref{Zorig})
the elasticity 
term $\int_0^t d\tau \frac{1}{4}(\frac{dx}{d\tau})^2$ is replaced by $ \int_0^t d\tau \frac{\kappa}{4}(\frac{dx}{d\tau})^2$ in Eq. (\ref{Zorig}) to study the effect of elasticity $\kappa$, the above result is trivially changed into 
$\overline{p^{(2)}_\eta(x;y|t)} = \frac{\kappa}{2 {\sf T} t}$. 

Let us now consider a discrete DP model 
on a lattice, e.g. as the one defined in the text. This model does not satisfy exact STS anymore.
However it can usually be described by an effective elastic constant $\kappa_{eff}(T)$
defined from the curvature around a minimum of the average free energy with respect to one end-point position. It may, in general depend on the temperature (defined for the lattice model). 
Hence we can conjecture that the large time limit of the observable $\overline{\hat p}$ defined
in the text will be $\overline{\hat p} \sim \frac{4 \kappa_{eff}(T)}{\hat t}$ (such that
for the discrete model defined in the text $\kappa_{eff}(T \to \infty)=1$).

\section{Representations of the symmetric group}

The eigenfunctions of $H_\partN$ in Eq.\eqref{LLHam} are contained in
$\mathcal{L}^2([0,L]^{\partN})$. Because of the integrability of the model,
the eigenfunctions can be written as linear combination of plane-waves in each sector $x_{Q_1} \leq \ldots \leq x_{Q_n}$
as in \eqref{BAwave}. Since $H_n$ is symmetric under
the exchange of coordinates, they can be classified according to the representations of the symmetric group.
For a fixed $P$, the vector of components $\pmatr{Q}{P}$ belongs to a vector space $V$ of dimension $\partN!$.
In this space it is naturally defined a representation 
of the symmetric group $\mathcal{S}_\partN$, called the regular representation, where the action of a permutation $T \in \mathcal{S}_\partN$ is simply the left multiplication
and the matrix representation is $\mathcal{L}(T)_{Q,Q'} = \delta_{T^{-1}Q, Q'}$. In a similar way, it is defined the dual representation based on the right 
multiplication $\mathcal{R}(T)_{Q,Q'} = \delta_{QT^{-1},Q'}$, 
associated with the exchange of two coordinates (see below). 
Note that $\mathcal{L}(T_1) \mathcal{L}(T_2) = \mathcal{L}(T_1 T_2)$ while
$\mathcal{R}(T_1) \mathcal{R}(T_2) = \mathcal{R}(T_2 T_1)$; moreover
$\mathcal{L}(T_1) 
\mathcal{R}(T_2) = 
\mathcal{R}(T_2) \mathcal{L}(T_1)$.

Moreover, the requirements for \eqref{BAwave} to be an eigenstates imposes that \cite{yang1967some}
\begin{equation}
 \label{relationAP}
     A_Q^{\sigma_i P} = \sum_{Q' \in \mathcal{S}_\partN } [Y^{\sigma_i}_{P_{i} P_{i+1}}]_{Q,Q'} A_{Q'}^{P}
 \end{equation}
where $\sigma_i$ is the transposition permutation exchanging $i,i+1$ and 
$Y^{\sigma_i}_{ab} = f(\mu_{ba}) \mathcal{L}(\sigma_i) + (f(\mu_{ba}) -1) \mathcal{L}(\mathbf{1})$
with $\mathbf{1}$ the identical permutation. 
This ensures that all the vectors $\pmatr{Q}{P}$
for different $P$ can be chosen inside an
irreducible representation 
of $\mathcal{S}_\partN$, which are in one-to-one correspondence with Young diagrams \cite{fulton1991representation}. 
It is useful to summarize the construction.
Given a standard tableau $\tabl$,
we associate two subgroups of $\mathcal{S}_\partN$: $\mathcal{A}(\tabl)$ and $\mathcal{B}(\tabl)$ defined respectively 
as the two subgroups that preserve respectively rows and columns, i.e. 
$\mathcal{A}(\tabl)$ are all permutations within rows, and 
$\mathcal{B}(\tabl)$ within columns.
Two elements of the group algebra of $\mathcal{S}_\partN$ are then defined as 
\begin{equation}
\label{abdef}
 a(\tabl) = \sum_{P \in \mathcal{A}(\tabl)} P \; , \quad b(\tabl) = \sum_{P \in \mathcal{B}(\tabl)} (-1)^{\sigma_P} P
\end{equation}
from which we define the Young symmetrizer 
\begin{equation}
\label{youngsymm}
c(\tabl) = a(\tabl) b(\tabl) = \sum_{P \in \mathcal{S}_{\partN}} c_P P\;. 
\end{equation}
The coefficients $c_P$ are integer numbers obtained from \eqref{abdef}. 
The Young symmetrizer projects onto $V_\tabl = \mathcal{R}(c(\tabl)) V$, an irreducible representation 
under the left-action $\mathcal{L}(\mathcal{S}_\partN) V_\tabl$, where $\mathcal{R}$ has been extended on the group algebra by linearity 
e.g. $\mathcal{R}(a(\tabl)) = \sum_{P \in \mathcal{A}(\tabl)} \mathcal{R}(P)$. 
The wave-functions built using the vectors $\pmatr{Q}{P}$ in $V_\tabl$ are anti-symmetric on the variables inside each column. Indeed for any permutation $T \in \mathcal{S}_{\partN}$:
\begin{equation}
 \label{wavefunctionsym}
  \psi_\mu(T\xv) = \sum_{P,Q \in \mathcal{S}_{\partN}} \vartheta_Q(\xv) \pmatr{Q T^{-1}}{P} \exp\bigl[i \sum_i x_{Q_i} \mu_{P_i}\bigr] 
  \end{equation}
using  that $(Q T {\bf x})_j = (T {\bf x})_{Q_j} = x_{T_{Q_j}}$ and performing the relabeling
$Q \to Q T^{-1}$. Now, for $T \in \mathcal{B}(\tabl)$ one finds that:
\begin{equation}
 \label{wavefunctionsym2}
  \psi_\mu(T\xv) =  (-1)^{\sigma_T} \sum_{P,Q \in \mathcal{S}_{\partN}} \vartheta_Q(\xv) \pmatr{Q}{P} \exp\bigl[i \sum_i x_{Q_i} \mu_{P_i}\bigr]
\end{equation}
In the last equality, we used that
for any vector $v \in V_\tabl$, $v = \mathcal{R}(b(\tabl)) v'$ for some $v' \in V$, and therefore
$\mathcal{R}(T) v = \mathcal{R}(b(\tabl) T) v' = \sum_{P \in  \mathcal{B}(\tabl)} (-1)^{\sigma_P} 
\mathcal{R}(P T) v' = (-1)^{\sigma_T} \mathcal{R}(b(\tabl)) v' = (-1)^{\sigma_T} v$.

An alternative path is to build the irreducible representations of $S_\partN$ using the Hilbert space $G$
of $\partN$ spin $1/2$. The action $\mathcal{G}(T)$ of a permutation $T$ is defined as the permutation of the spins 
\begin{equation}
 \label{spinaction}
 \mathcal{G}(T)\kket{i_1\ldots i_\partN} = \kket{i_{T_1}\ldots i_{T_\partN}}
\end{equation}
where $i_k \in \{\uparrow, \downarrow\}$, the two eigenstates of the $k$-th spin along the $z$-direction. The sub-space $G_\diagr$ of highest weights, i.e. annihilated by $S^+ = \sum_k s^+_k$, with fixed total magnetization $S^z = \sum_k s^z_k =  (\partN - 2 \partM)/2$, defines the irreducible representation
corresponding to the two-rows diagram $\diagr = (\partN - \partM, \partM)$. 
This procedure can be extended to the general case and allows deriving the equations for the rapidities $\pv$, together with a
hierarchy of auxiliary rapidities for each row of $\diagr$, called nested-Bethe-Ansatz equations \cite{sutherland1968further}. For the two-rows diagrams, 
we get \eqref{NBAeq} in the text. 
Although these equations only depend on the diagram $\diagr$, 
wave-functions depend on 
the tableau $\tabl$. For instance, for $\partN = 3$ and $\partM = 1$, we have two possible tableaux
\begin{equation}
\label{exampletableau}
\Yvcentermath1
 (2,1)_1 = \young(12,3) \;,  \qquad (2,1)_2 = \young(13,2) \;.
\end{equation}
All the possible standard tableaux correspond to different copies of a given irreducible representation
inside the regular one and therefore to different multiplets of wave-functions. 
For a general two-row diagram the dimension of the irreducible representation is 
\begin{equation}
d_\xi=d_{(n-m,m)}={n \choose{m}} \left(\frac{n-2m +1}{n-m+1} \right)
 \end{equation}
and its multiplicity inside the regular representation
is also equal to $d_\xi$ \cite{fulton1991representation}.
The spin-representation can be mapped into each of these copies by a linear mapping $\Psi: V \to G$, with components
$\Psi(v) = \sum_Q v_Q \kket{\Psi_Q}$, for any vector $v \in V$ of components $v_Q$. 
A complete characterization of this mapping is obtained requiring that, for each choice $\tabl$ for the same $\diagr$,
\begin{itemize}
 \item $\Psi(V_\tabl) \equiv G_{\diagr}$, unitarily so that, for any $w \in V_{\tabl}$
\begin{equation}
\label{normconserve}
 || w ||^2 \equiv \sum_Q |w_Q|^2 = || \Psi(w) ||^2 
\end{equation}
 \item for any $w \in V_{\tabl}$:  
 \begin{equation}
 \label{reprcommut}
\Psi(\mathcal{L}(T) w) = \mathcal{G}(T) \Psi(w)   
 \end{equation}
\item we fix the action of $\Psi$ on the space orthogonal to $V_{\tabl}$ by
$\sum_Q \kket{\Psi_Q} \bbra{\Psi_Q} = \Pi_\xi$, with $\Pi_\xi$ the projector of $G$ onto $G_\xi$.
 \end{itemize}
With these definitions, we take
\begin{equation}
 \label{APQSpinChain0}
 \pmatr{Q}{P} = \pmatr{\text{\tiny bos}}{P}\bbrakket{\Psi_Q}{\omega(P\pv)}\;.
\end{equation}
In practice, the explicit form of $\Psi$ is not needed, since the differential operator $\mathcal{D}_\partM$ gives a non-vanishing result only
for one Young tableau $\tabl_0$, corresponding
to the filling described in \eqref{youngtabex}. 
In order to show this, we now give the explicit derivation of Eq.(12) in the main text.
The vector of components $v_Q^{(\partM)} = d_\partM(Q^{-1} \pv)$ belongs to $V_{\tabl_0}$
and we only need the image $\kket{\partM} \equiv \Psi(v^{(m)}) 
= \sum_Q d_\partM(Q^{-1} \pv) \kket{\Psi_Q}
$  of this vector in the spin representation $G_{\tabl_0}$. To determine it, we expand
it as
\begin{equation}
 \kket{\partM} = \sum_{a_1,\ldots,a_\partM=1}^\partN q_{a_1,\ldots,a_\partM}(\pv) \sigma_{-}^{a_1}\ldots \sigma_{-}^{a_m}\kket{+} 
\end{equation}
where $q_{a_1,\ldots,a_\partM}(\pv)$ is a symmetric tensor obtained as a linear combination of the $v_Q^{(\partM)}$. Its components are therefore 
homogeneous polynomials in the $\pv$ of degree $m$Because of \eqref{reprcommut}, under the mapping $\Psi$, a permutation of the rapidities $\pv$ 
corresponds to a permutation of spins, we deduce the expansion
\begin{equation}
q_{a_1,\ldots,a_\partM}(\pv) = \sum_{k=0}^m q_k^1(\pv) q_{m-k}^2 (\mu_{a_1},\ldots,\mu_{a_m})
\end{equation}
where $q_k^{1,2}$ are symmetric polynomials of degree $k$.
Moreover, from
$S^{+}\kket{m} = 0$, we have
\begin{equation}
 \label{highestweight}
 \sum_{\substack{a=1\\a\neq a_2,\ldots,a_\partM}}^\partN q_{a a_2\ldots a_\partM} = 0 \;.
\end{equation}
These conditions are sufficient to determine the state $\kket{\partM}$ 
and in particular for $\partM=1$, they lead to 
\begin{equation}
 q_{a} = \sqrt{Z} (\mu_{a} - \avemu) \;,
\end{equation}
where $\avemu=\frac{1}{n} \sum_{a=1}^n \mu_a$ and, since $\Psi$ satisfies \eqref{normconserve}, $Z$ can be fixed equating  
$\sum_Q | v_Q^{(\partM)} |^2 \equiv || v_Q^{(\partM)} ||^2 = \bbrakket{\partM}{\partM}$, recovering
\eqref{sumQ}.

\section{Nested Bethe ansatz approach}

\subsection{Norm of a state: general form}
The expression of the norm for a given set of $\pv$ and $\lv$ solutions of \eqref{NBAeq} on a circle of length $L$ 
takes the form 
\begin{equation}
||\psi_\mu||^2 \equiv \int_0^L |\psi_\mu(x)|^2 d\xv =
\sum_{P,P' \in \mathcal{S}_\partN} \left(\sum_Q \pmatr{Q}{P} (\pmatr{Q}{P'})^\ast\right) \int d\xv \, \theta(\xv)  
e^{i \sum_{j=1}^\partN (\mu_{P_j} - \mu_{P_j'}) x_j } 
\end{equation}
Now the sum over $Q$ can be performed employing \eqref{APQSpinChain}. Since 
$\sum_Q \kket{\Psi_Q} \bbra{\Psi_Q} =  \Pi_{\xi}$, it acts as the identity inside $G_\xi$ and we have 
\begin{equation} \label{norm1} 
||\psi_\mu||^2 = \sum_{P,P'} \int d\xv \, \theta(\xv)  
e^{i \sum_{j=1}^\partN (\mu_{P_j} - \mu_{P_j'}) x_j } A^P_{\text{\tiny bos}} A^{P'}_{\text{\tiny bos}} 
\bbrakket{\omega(P'\mu)}{\omega(P\mu)} \;.
\end{equation}
This last expression is the norm of a state of a two-components Lieb-Liniger model. This integral 
can be computed exactly in the framework of algebraic Bethe-ansatz \cite{pozsgay2012form}, leading to
\begin{equation}
 \label{pozsgaynorm}
 || \psi_\mu ||^2 = c^{m} B(\lv) \det \left(\begin{array}{cc} G_{\mu\mu} & G_{\mu\lambda} \\ G_{\lambda \mu} & G_{\lambda\lambda} 
\end{array}
\right)
\end{equation}
where the matrices $G_{\mu\mu}\in \partN\times\partN, G_{\mu\lambda}\in \partN\times\partM, G_{\lambda\lambda} \in \partM\times\partM$ are given as 
\begin{align}
 (G_{\mu\mu})_{jk} &= \delta_{jk}(L - \sum_{a=1}^{\partM} Q_{ja} 
 + \sum_{l=1}^{n} K_{jl}) - K_{jk} \\
 (G_{\mu\lambda})_{ja} &= (G_{\lambda\mu})_{aj} = Q_{ja} \\
 (G_{\lambda\lambda})_{ab} &= \delta_{ab} \left(\sum_{j=1}^{\partN} Q_{ja} - \sum_{a'=1}^{\partM} K^{(\lambda)}_{aa'} \right) + K^{(\lambda)}_{ab}
\end{align}
and we introduced the notation
\begin{equation} 
B(\lv) \equiv \prod_{1\leq a<b \leq \partM} \left(\frac{\lambda_{ab}^2 + c^2}{\lambda_{ab}^2}\right)
\end{equation}
\begin{equation}
Q_{ja} \equiv \frac{c}{(\frac{c}{2})^2 +(\mu_j - \lambda_a)^2}\;,\quad
 K_{jl} \equiv \frac{2c}{c^2 +(\mu_j - \mu_l)^2}\;, \quad
 K_{ab}^{(\lambda)} \equiv \frac{2c}{c^2 +(\lambda_a - \lambda_b)^2} \;.
\end{equation}

\subsection{String states and their norms}

Examination of the NBA equations, and consistency
with the nested contour integral method (see section below), indicate that at large 
$L$ the rapidities of the eigenstates 
$\pv$ are arranged in a set of strings, as is the case for bosons:
\begin{equation}
 \label{stringdeviations}
\mu_{j}^{a} = k_j + \frac{i c}{2} (m_j + 1 -2 a) + \delta_j^{a} 
\end{equation}
where $a=1,\ldots,m_j$, $j=1,\ldots, \partNS$ and $\partNS$ is the number of strings, 
while $m_j \geq 1$ is the size of the $j$-th string (called an $m_j$-string). 
For $m_j=1$, $\mu_j=k_k$ is real (called a $1$-string). 
The $\delta_j^a$
are the deviations from the string hypothesis, 
in general different from the bosonic ones, but that
we can still assume to be exponentially small for large $L$.

An important difference with the bosonic case is that now {\it some of
the strings are missing}. Let us call $\Pi^j_m$ the factor containing $\lambda_a$ 
in \eqref{NBAeqmu}. In the bosonic case $m=0$ this factor is set to
unity and one recovers the usual BA equation: from its poles and zeros one
sees that the strings \eqref{stringdeviations} are the solutions. 
For $m \geq 1$ the factor $\Pi^j_m$, upon substitution of 
$\lambda_a$ from solving \eqref{NBAeqla}, will sometimes cancel
some of these poles and zeros and some strings will be missing.
Let us show two explicit examples for $m=1$, in which case
there is a single $\lambda_a=\lambda$. 

Consider $\partN=2$, $\partM=1$. Then the solution of \eqref{NBAeqla}
is $\lambda_1 \equiv \lambda=\frac{1}{2} (\mu_1+\mu_2)$ and \eqref{NBAeqmu} 
becomes simply $e^{i L \mu_1}=e^{i L \mu_2}=1$.
The $2$-string states are missing, the only states contain two $1$-strings,
with, in fact free particle momenta quantization. This is expected from
the antisymmetry of the corresponding Young diagram
which make the two particles fermions unaffected by the $\delta$ interaction.

Consider now $n=3$, $m=1$. One looks, with no loss of generality, for
solutions such that $\mu_1+\mu_2+\mu_3=0$, and such that 
$\mu_3$ is real and $\mu_2=\mu_1^*$. From considering the
modulus of \eqref{NBAeqmu} for $j=3$ one finds that $\lambda$ must be real.
On the other hand there are now two solutions to \eqref{NBAeqla} 
$\lambda= \pm \frac{1}{2 \sqrt{3}} \sqrt{ c^2 + 4 (\mu_1+\mu_2)^2 - 4 \mu_1 \mu_2}$.
At large $L$ one then finds two types of solutions of \eqref{NBAeq}: either (i) three 1-strings
with all $\mu_j$ real and the usual quasi-free quantization conditions. Or 
(ii) $\mu_{1,2}=\frac{1}{2} (k \pm i s)$, $s >0$, and $\mu_3=-k$. 
The r.h.s. of \eqref{NBAeqmu} for $j=1$ vanishes at large $L$, hence
we look for zeros of the l.h.s. There are two such zero es: one for $s=\bar c$,
which leads to the $2$-string plus $1$-string state, hence allowed here.
The other zero requires $s=2 \bar c$ and $k=0$ which is excluded by
the condition that $\lambda$ is real (which implies $s^2 < c^2 + 3 k^2$).
Hence the $3$-string state is missing. 

We will continue by assuming that in all cases this structure remains, and that 
for allowed string configurations, the $\lambda_a$ do not introduce additional singularities
in the modified Gaudin matrix displayed above.
Then following \cite{calabrese2007dynamics}, we deduce the following large $L$ limit for $||\psi_\mu||^2$ 
$$ || \psi_\mu||^2 = \left[\prod_j \prod_{a=1}^{m_j -1} \frac{1}{\delta^a_j - \delta^{a+1}_j}\right]  
\left[\prod_{j=1}^{\partNS} m_j\right] c^\partM L^{\partNS} B(\lv)  \det G_{\lambda\lambda} $$ 
Now consider the factor 
$$(\Omega_{\pv}^0)^2 = \prod_{i<j} \frac{(\mu_i - \mu_j)^2}{(\mu_i - \mu_j)^2 + c^2} = \prod_{i\neq j} \frac{\mu_i - \mu_j}{\mu_i - \mu_j - i c} $$
and again we insert the rapidities organized in strings as in \eqref{stringdeviations}. To do so we rewrite
$$ (\Omega_{\pv}^0)^2 = \prod_{(j,a) \neq (j',a')} \frac{\mu_j^a - \mu_{j'}^{a'}}{\mu_j^a - \mu_{j'}^{a'} - i c}
$$
This product splits into two contributions. The products inside same string $j$ that we label $O_j$ and the products from two different strings $j\neq j'$ that we label $O_{jj'}$. 
These two contributions can be written as
\begin{align}
O_j &= \prod_{a\neq a'} \frac{\mu_j^a - \mu_{j}^{a'}}{\mu_j^a - \mu_{j}^{a'} - i c} =
\left(\prod_{a} \frac{1}{i(\delta_j^{a+1} - \delta_{j}^{a})} \right) 
\frac{\prod_{a\neq a'=1}^{m_j}(ic(a-a'))}{\prod_{\substack{a\neq a'\\a\neq a'+1}}^{m_j}(ic(a-a'-1))}=
\left(\prod_{a} \frac{1}{\delta_j^{a} - \delta_{j}^{a+1}} \right) \frac{c^{m_j-1}}{m_j}\\
O_{jj'} &= \prod_{a=1}^{m_j} \prod_{a'=1}^{m_{j'}} \frac{k_j - k_{j'} + \frac{i c (m_j-m_{j'})}{2} + i c (a' - a)}
 {k_j - k_{j'} + \frac{i c (m_j-m_{j'})}{2} + i c (a' - a - 1)} = \prod_{a'=1}^{m_{j'}} \frac{k_j - k_{j'} + \frac{i c (m_j - m_{j'})}{2} + i c (a'-1)} 
 {k_j - k_{j'} - \frac{i c (m_j+m_{j'})}{2} + i c (a'-1)} 
\end{align}
So that we have
\begin{equation} O_{j j'} O_{j' j} = \prod_{a=1}^{m_j} 
\left(\frac{k_j - k_{j'} + \frac{i c (m_j + m_{j'})}{2} - i c a} 
 {k_j - k_{j'} + \frac{i c (m_j-m_{j'})}{2} - i c a} 
\right)\left( \frac{k_j - k_{j'} + \frac{i c (m_j - m_{j'})}{2} - i c (a-1)} 
 {k_j - k_{j'} + \frac{i c (m_j+m_{j'})}{2} - i c (a-1)} \right)\\
 = \left(\frac{(k_j - k_{j'})^2 + \frac{c^2 (m_j - m_{j'})^2}{4}}{(k_j - k_{j'})^2 + \frac{c^2 (m_j + m_{j'})^2}{4}}\right)
\end{equation}
and finally 
\begin{equation}
( \Omega_{\pv} ^0)^2= \left(\prod_j O_j \right) \left(\prod_{j<j'} O_{jj'}O_{j'j}\right) =  c^{N-\partNS}\left(\prod_{j} \frac{1}{m_j}\prod_{a} \frac{1}{\delta_j^{a} - \delta_{j}^{a+1}} \right)\Phi(\kv, \mv) \;.
\end{equation}
It follows that the following ratio is finite
\begin{equation}
\label{normratio}
 \frac{(\Omega_{\pv}^0)^2}{||\psi_\mu||^2}= \frac{\bar c^{n-\partNS - m} \Phi(\kv, \mv)  \prod_{j} \frac{1}{m_j^2}}{L^{\partNS} B(\lv)  \det G_{\lambda\lambda} }
\end{equation}
Note that, in the bosonic case 
$m=0$, in order to enforce the condition $\sum_Q \kket{\Psi_Q}\bbra{\Psi_Q} = \Pi_\xi$,
one has to choose $\kket{\Psi_Q} = \frac{\kket{\uparrow\ldots \uparrow}}{\sqrt{\partN!}}$.
Inserting this expression in \eqref{APQSpinChain0}, we see that with these conventions
$\pmatr{P}{Q} = A^P_{\text{\tiny bos}}/\sqrt{n!}$. This is the origin of the missing $\partN!$ factor
in \eqref{normratio}, compared for instance to \cite{calabrese2010free}.

It is easy to check the formula \eqref{normratio} in the case of $n_s=n$ $1$-strings,
i.e. all $\mu_j$ real. Then in (\ref{norm1}) one finds that at large $L$ the term $P'=P$ 
gives the leading contribution, hence:
\bea
 ||\psi_\mu||^2 = \frac{L^n}{n!} \sum_{P,Q} |A^P_Q|^2 = 
 L^n \sum_P |A^P_{\text{\tiny bos}}|^2 || \omega(P\mu) ||^2
\eea 
For simplicity let us consider $m=1$. Then $\kket{\omega(\pv)} = \sum_{a= 1}^\partN \kappa_{a} ( \lambda | \pv ) 
\sigma_{-}^{a_1} \kket{+}$. Since all $\mu_j$ are real (as well as $\lambda$, as discussed above)
we find:
\bea
|| \omega(P\mu) ||^2 = \sum_{a=1}^n |\kappa_a(\lambda | \pv  )|^2 = 
\sum_{a=1}^n \frac{c^2}{(\lambda- \mu_a)^2 + c^2/4}  
\eea 
in agreement with (\ref{normratio}) and the formula given in the text (for $m_j=1$ and $n_s=n$).

\subsection{Explicit derivation of $\partM=1$ case: re-summation over the spin-rapidities}
The action of the differential operator $\mathcal{D}_1$ 
on the wave-function in \eqref{BAwave}, after averaging on the different sectors $x_{Q_1} < x_{Q_2} \ldots $,
can be written explicitly using
\eqref{APQSpinChain} in the text and leads to
\begin{equation}
 \label{BAwavederiv}
  \mathcal{D}_1\psi_{\pv,\lambda}(\zero)  = \frac{i}{\partN!}\sum_{P} \pmatr{\text{\tiny bos}}{P}  \sum_Q \underbrace{\frac{\mu_{(Q^{-1}P)_1}-\mu_{(Q^{-1}P)_2}}{\sqrt{2}}}_{d_1(Q^{-1} P \pv)} \bbrakket{\Psi_Q}{\omega(P\mu)} \;.
\end{equation}
The sum over $Q$ can be performed using \eqref{sumQ}, leading to
\begin{equation}
 \label{BAwavederiv1}
  \mathcal{D}\psi_{\pv,\lambda}(\zero)  = i \sqrt Z \sum_{P} \pmatr{\text{\tiny bos}}{P}  \sum_{a=1}^n (\mu_{P_a}-\hat \mu) \kappa_a(\lambda, P \pv) =
  i \sqrt{Z} \partN! \left[\prod_{1 \leq j<l \leq n}\frac{\mu_{l} - \mu_{j}}{\mu_{l} - \mu_{j} - i c}\right] \mathcal{F}(\lambda, \pv)
\end{equation}
where $\mathcal{F}(\lambda, \pv)$ takes the explicit form
\begin{equation}
 \label{Fexplicit}
 \mathcal{F}(\lambda, \pv) = 
 \frac{ic}{\partN!}\sum_P \left[\prod_{1 \leq j<l \leq n} \frac{\mu_{P_l} - \mu_{P_j} - i c}{\mu_{P_l} - \mu_{P_j}}\right]
\sum_{a=1}^n (\mu_{P_a}-\hat \mu) \frac{\prod_{d=a+1}^n (\lambda - \mu_{P_d} - i c /2)}{\prod_{d=a}^n (\lambda - \mu_{P_d} + i c /2)} \;.
\end{equation}
Therefore, the norm square is equal to
\begin{equation}
 \label{BAwavederivsq}
  |\mathcal{D}\psi_{\pv,\lambda}(\zero)|^2  = \partN(\partN-2)! |\Omega_{\pv}|^2
  |\mathcal{F}(\lambda, \pv)|^2 \;.
\end{equation}
By taking into account the norm of the wave-function and \eqref{normratio}, 
we see that the ratio $|\mathcal{D}\psi_\mu(\xv)|^2/||\psi_\mu||^2$ remains finite
in the $L\to\infty$ limit. 
In this limit the string momenta $k_j$ become arbitrary real number
but a multiplet of wave-functions 
is obtained in correspondence of the set of $\{\lambda^\ast\}$, solutions of \eqref{NBAeqla}. Specializing \eqref{normratio} to $m=1$
we obtain
\begin{equation}
 \label{sumlambda}
 \sum_{\lambda^\ast} \frac{|\mathcal{D}\psi_{\pv,\lambda^\ast}(\zero)|^2}{|| \psi_{\pv,\lambda^\ast}||^2} 
 = \frac{\partN! \Phi(\kv, \mv)  \prod_{j} \frac{1}{m_j^2}}{c^{\partNS - \partN}L^{\partNS}}
\underbrace{\sum_{\lambda^\ast} 
\frac{i c^{-1} P_-(\lambda^\ast) P_+(\lambda^\ast) |\mathcal{F}(\lambda^\ast, \pv)|^2}
{(N-1)(P_+(\lambda^\ast) P_-'(\lambda^\ast) - P_+'(\lambda^\ast) P_-(\lambda^\ast) )}}_{\QQ_{\partN,1}(\pv)}
 \end{equation}
where we used that for $m=1$, we have the equality
$$\det G_{\lambda\lambda} = \sum_a \frac{c}{(\lambda - \mu_a)^2 + c^2/4} = -i\partial_\lambda \log P_-(\lambda)/P_+(\lambda) \;.$$
and we have $B(\lambda)=1$ for $m=1$. 
The sum over the solutions of \eqref{NBAeqla} can then be replaced by
$$ \QQ_{\partN,1} = \frac{1}{\partN -1}\oint \frac{dz}{2\pi c} \frac{P_-(z) P_+(z) |\mathcal{F}(z, \pv)|^2}
{P_+(z) P_-'(z) - P_+'(z) P_-(z) } \partial_z \log\left( \frac{P_-(z)}{P_+(z)}- 1\right) = 
\frac{1}{\partN -1}\oint \frac{dz}{2\pi c} \frac{|\mathcal{F}(z, \pv)|^2 P_-(z)}{P_-(z) - P_+(z)}
$$
where the contour encloses only the roots $\lambda^\ast$ and no other singularity of the integrand. 
The integral can also be computed by considering the poles outside the contour,
i.e. the poles of $|\mathcal{F}(z,\pv)|^2$, at the zeros of $P_+(z)$: $z_b = \mu_b- i c /2$. 
We have therefore calculating the residue
\begin{equation}
 \label{sumoverlambda1}
 \QQ_{\partN,1}	(\pv) = 
\frac{1}{\partN - 1}
\sym_{\pv} \left[\left(\prod_{j<l} \frac{\mu_{l} - \mu_{j} - i c}{\mu_{l} - \mu_{j}}\right) 
 g(\pv) 
 \right]
 \end{equation}
 where we set 
 \begin{equation}
 \label{tildeg}
g(\pv)  =  -\frac{i c}{N!} \sum_Q \left[\prod_{j<l} \frac{\mu_{Q_l} - \mu_{Q_j} + i c}{\mu_{Q_l} - \mu_{Q_j}}\right] \\ 
 \sum_{\substack{a,a' \\ b\geq a}}
 (\mu_{a}-\avemu)(\mu_{Q_{a'}}-\avemu)\frac{\prod_{d>a} (\mu_b - \mu_{d} - i c)}{\prod_{\substack{d\geq a\\d\neq b}} (\mu_b - \mu_{d})}
\frac{\prod_{d'>a'} (\mu_b - \mu_{Q_d})}{\prod_{\substack{d'\geq a'}} (\mu_b - \mu_{Q_d}- ic)}
 \end{equation}
 where in taking the complex modulus square we have used the fact that
 $\pv$ and its complex conjugate are identical up to a permutation.
  
In order to make more explicit the expression for $\QQ_{\partN,1}(\pv)$, 
we start noticing that $g(\pv)$ is a polynomial in the $\mu$'s, as $g(\pv)$ has no singularities. 
Indeed
\begin{itemize}
 \item For $\mu_\alpha = \mu_\beta + \epsilon$, for arbitrary $\alpha,\beta$ we can 
 have singular terms of order up to $\epsilon^{-2}$.
For $\alpha,\beta \geq a$, we can possibly have a term of order $\epsilon^{-2}$ but then the terms with $b=\alpha$ and $b = \beta$ cancel each other. 
Singularity of order $\epsilon^{-1}$ come from the first
product term in \eqref{tildeg}. But this cancels out between the permutations $Q$ and $Q \sigma_{\alpha\beta}$, where $\sigma_{\alpha\beta}$ 
is the transposition exchanging indexes $\alpha,\beta$.
 \item For $\mu_\alpha = \mu_\beta -
 i c + \epsilon$, we can have a singularity of order $\epsilon^{-1}$, coming from the last product in \eqref{tildeg}, when $Q_d = \alpha$ and $b=\beta$. Therefore
 for this to be there, we need $(Q^{-1})_\beta  > (Q^{-1})_\alpha$ (otherwise the zero of the first term would cancel the singularity) and 
 $(Q^{-1})_\beta \leq a'$, which shows that the condition is never realized since $(Q^{-1})_\alpha \geq a'$. 
\end{itemize}
Now, with a similar procedure it is easy to see that also $\QQ_{\partN,1}(\pv)$ has to be a polynomial. 
Moreover it is symmetric in the $\mu$'s,
and homogeneous in the $\mu$'s and $c$, and is not changed under the transformation $\mu_i \to \mu_i + z$ for any $z$. 
Therefore $$ \QQ_{\partN,1}(\pv) = \frac{1}{\partN-1}(G^{(2)}(\pv) + c G^{(1)}(\pv) + c^2 G^{(0)})$$
where $G^{(k)}(\pv)$ is homogeneous of degree $k$ in the $\mu$'s . Now $G^{(2)}(\pv)$ can be computed setting $c\to 0$. In \eqref{tildeg},
one sees that the limit $c\to 0$ imposes $a = Q_{a'} = b$ and therefore	
$$ G^{(2)}(\pv) = \sum_a (\mu_a - \hat \mu)^2 = \frac{1}{\partN} \sum_{i<j} (\mu_i - \mu_j)^2 $$ 
Moreover $G^{(1)}(\pv) = 0$, being the only possible symmetric polynomial, function of differences, of degree $1$. 
Since $g(\pv) = 0$ for $\mu_j = -i j c$, we deduce that $G^{(0)} = \partN(\partN^2-1)/12$. Finally we get
$$ \sum_{\lambda^\ast} \frac{|\mathcal{D} \psi_{\pv, \lambda}(\zero)|^2}{||\psi_{\pv, \lambda}||^2} =
\frac{(\partN-2)! 
\Phi(\kv, \mv)  \prod_{j} \frac{1}{m_j^2}}{\bar c^{\partNS - \partN}L^{\partNS}} \left.
\left(\sum_{i<j} (\mu_i - \mu_j)^2 + \frac{\partN^2(\partN^2-1)c^2}{12}\right)\right|_{\pv = \pv(\kv,\mv)}
$$
where in the last expression the $\mu$'s have to be organized in the string ansatz.

\section{General initial conditions: residue expansion of contour-integral}
\begin{figure}[t]
\begin{minipage}{0.495\linewidth}
\centering
\includegraphics[width=0.9\columnwidth]{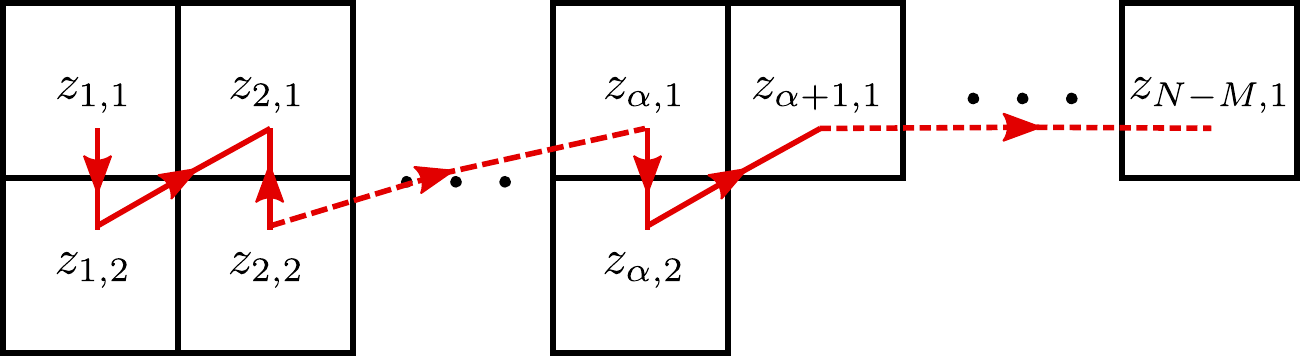}
  \end{minipage}
\begin{minipage}{0.495\linewidth}
\centering
\includegraphics[width=0.7\columnwidth]{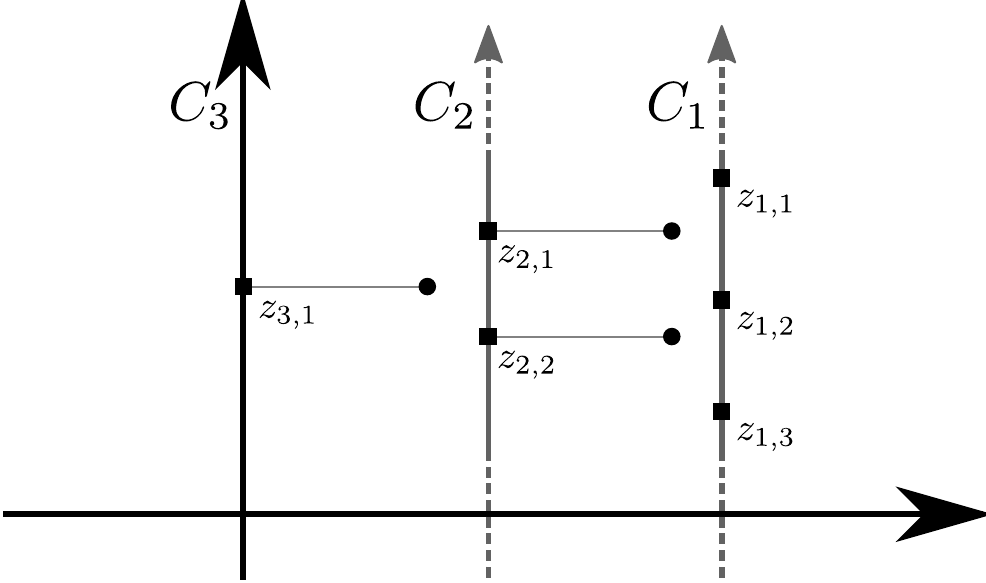}
  \end{minipage}
\caption{\textbf{Left --} Sketch of the top-bottom/left-right filling of variables for a two-rows Young diagram. The ordering of the variables is represented
  by the arrows. \textbf{Right --} Structure of the poles (black disks) at $z_{\alpha,i}+1$ with $\alpha > 1$ 
  when integrating the variables $z_{1,j}$, $j=1,2,3$ with $\xi = (321)$.
  \label{fig:diagramSketch}}
  \end{figure}

Suppose now we are dealing with an initial condition, whose properties under exchange can be
encoded in the Young diagram $\xi$ filled with coordinates $\xv$: we require the wave-functions to be
antisymmetric when two coordinates in a
column of $\xi$ are exchanged. In \cite{borodin2014macdonald} 
the disorder average of products of partition sums of groups of non-crossing
directed paths was considered. 
Each such product can be associated to a Young diagram $\xi$
and involves $n_C(\xi)$ groups, each group containing 
$\xi^{\alpha}$ paths mutually non-crossing within each group and
with coinciding starting points. Based on a similar theorem obtained there for
the semi-discrete directed polymer, it was conjectured (remark 5.25 there) that for the continuum
model the global partition function 
$\mathcal{Z}_{\xi}(T) = \overline{\prod_{\alpha=1}^{\nc{\xi}} {\cal Z}_{\xi^\alpha}(T,X_\alpha)}$ 
(see notations there) can be written as a multiple contour integral
\begin{equation}
 \label{Zborodin}
 \mathcal{Z}_{\xi}(T) = \prod_{\alpha=1}^{\nc{\xi}} \frac{1}{(2\pi i)^{\xi^\alpha} \xi^\alpha!} \prod_{j=1}^{\xi^\alpha} \int 
dz_{\alpha,j}  \prod_{1 \leq  \alpha < \beta \leq \nc{\xi}} \left( \prod_{j=1}^{\xi^\alpha} \prod_{j=1}^{\xi^\beta} \frac{z_{\alpha,i} - z_{\beta,j}}{z_{\alpha,i} - z_{\beta,j} - 1}
 \right) 
 \times \prod_{\alpha=1}^{\nc{\xi}} \left( \prod_{i \neq j}^{\xi^\alpha} (z_{\alpha,i} - z_{\alpha,j}) 
 \right) F(\{z\})\;.
\end{equation}
where the integration domain for the variable $z_{\alpha,j}$ is chosen along $C_\alpha + i \mathbb{R}$, i.e.
parallel to the imaginary axis, with $C_1 > C_2 + 1 > \ldots > C_k + (k-1)$. 
Here we set
\begin{equation}
F(\{z\}) =  \prod_{\alpha=1}^{\nc{\xi}} \prod_{j=1}^{\xi^\alpha} e^{\frac{T}{2} z_{\alpha,j}^2  + X_\alpha z_{\alpha,j}} 
\end{equation}
and in particular we assume it to be is an entire function of the variables $\{z\}$. 
For 
\begin{equation}
\label{rescfrombor}
X_\alpha = 0,\quad T = 2 \cbar^2 t,\quad \cbar^{2\partM + n} \mathcal{Z}_\xi (T) \to \Theta_{\partN, \partM}
\end{equation} 
we recover \eqref{BorodinContour}
when $\xi = (\partN-\partM, \partM)$, i.e. a two-rows diagram
with $\xi^1 = \xi^2 = \ldots = \xi^{\partN-\partM} = 2, \xi^{\partN-\partM+1}=1,\ldots, \xi^{\partN} = 1$. 
To see this it is enough to perform the change of variables $z_{\alpha,j} \to i z_A / \cbar $, where $A=1,\ldots, \partN$, with 
the ordering depicted in Fig.~\ref{fig:diagramSketch} (Left). Note that contours change
as given in the text, where we have also distinguished contours for variables with the 
same index $\alpha$ (which is immaterial since no poles are encountered when bringing 
them together). 

We will now show that, as for the bosonic case \cite{borodin2014macdonald} where all $\xi^\alpha=1$, 
this expression is equivalent to the spectral expansion \eqref{pGenAve} in the text, into string solutions 
of the Bethe Ansatz equations. We use the notations of \footnote{Borodin private communication.}
and extend the method to the case of a general symmetry. 
Starting from $z_{1,1}$ and the contour $C_1$, we recursively move all the contours to the leftmost $C_{\nc{\xi}}$. 
While displacing the contours, all the residues of the first product in \eqref{Zborodin} have to be collected: 
they produce a hierarchy of additional terms with fewer integration variables $z_{\alpha_k,i_k}$, which are the ending points of 
a sequence of residues $z_{\alpha_1,i_1} = z_{\alpha_2, i_2}+1 = z_{\alpha_3, i_3}+2 = \ldots = z_{\alpha_k, i_k} + k-1$.
Each of these terms is characterized by the length $\chi_\alpha = k$ of these sequences, that we arrange
in another diagram of size $\partN$:
$\chi = (\chi_1\chi_2\ldots)$. 
A filling $\phi$ of the diagram $\chi$ corresponds to assign the variables $z_{\alpha,i}$ to each box thus
fixing the sequence of residues. For instance 
for $\xi = (42)$ and $\chi = (321)$ one possible filling is:
\begin{equation}
\label{examplefilling}
\raisebox{-15mm}{ \includegraphics[keepaspectratio=true,height=0.15\columnwidth]{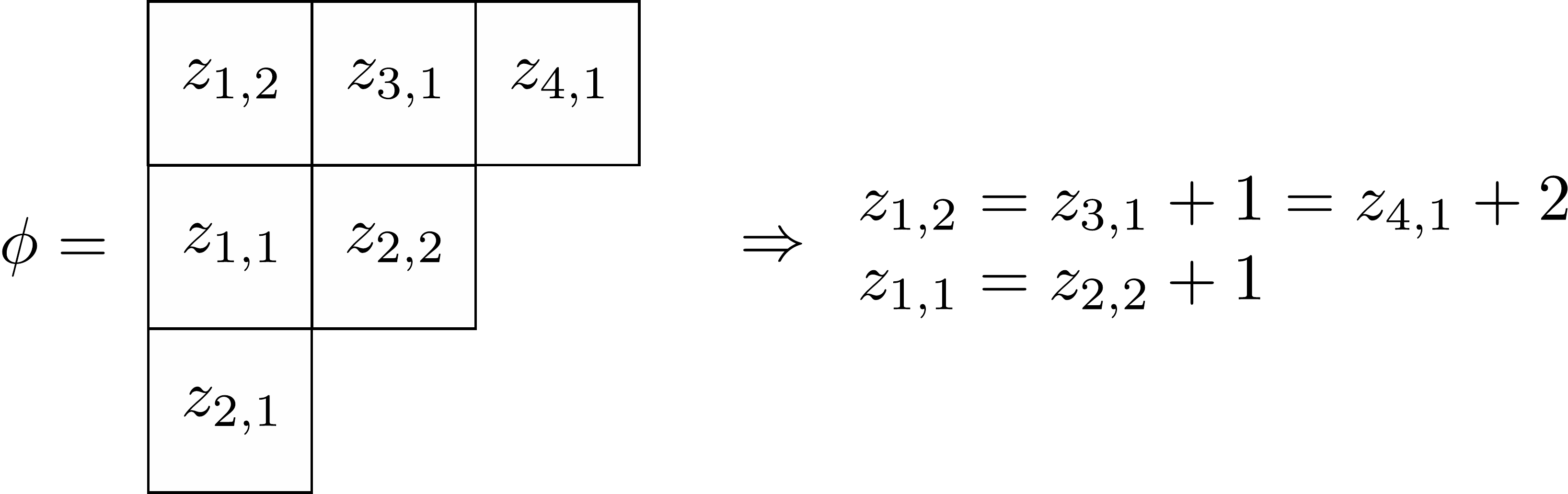}}
\end{equation}
and in general we indicate as $z_{\phi_{i,j}}$ the variable in the box at row $i$ and column $j$ 
of the filling $\phi$. 
In order to have a non-vanishing residue, each index $\alpha$ has to grow strictly along each row.
This restricts the possible diagrams $\chi$ to those satisfying $\chi \prec \xi$, 
meaning that either $\xi \equiv \chi$
or $\xi_i > \chi_i$, where $i$ is the smallest index where they differ. 
Then Eq.~\eqref{Zborodin} is expanded as a sum over all the elements of the set of allowed fillings 
$\Phi(\chi)$ of the diagrams $\chi \prec \xi$
\begin{multline}
 \label{step1}
 \mathcal{Z}_{\xi}(T) = \prod_{\alpha=1}^{\nc{\xi}} \frac{1}{\xi^\alpha!}\sum_{\stackrel{\chi \prec \xi}{\chi = 1^{a_1} 2^{a_2} \ldots}} \frac{1}{a_1! a_2!\cdots} \sum_{\phi \in \Phi(\chi)} \int \cdots \int 
 \prod_{p=1}^{\nr{\chi}} {\frac{dz_{\phi_{p,\chi_p}}}{2 \pi i}}\\
 \times \Res_{\phi} \left[\left(\prod_{\alpha<\beta} \prod_{i,j} \frac{z_{\alpha,i} - z_{\beta,j}}{z_{\alpha,i} - z_{\beta,j} - 1}
 \right) \left( \prod_{\alpha=1}^{\nc{\xi}} \prod_{i \neq j}^{\xi^\alpha} (z_{\alpha,i} - z_{\alpha,j}) 
 \right) F(\{z\})\right] \;.
\end{multline}
The multiplicities $a_k$ are the number of indexes $\alpha$ such that  $\chi_\alpha = k$ and avoid 
the overcounting when a diagram $\chi$ has multiple rows of the same length, i.e. multiple strings of the same length.
The expression $\Res_\phi[G(\{z\})]$ entails the iterative computation of all the residues at $z_{\phi_{k,l}} = z_{\phi_{k,l+1}}+1$
for $k=1,\ldots,\nr{\chi}$ and $l=1,\ldots, \chi_{k}-1$,
obtaining finally a function of the rightmost variables $z_{\phi_{p,\chi_p}}$ in each row of $\phi$.
Now we consider the relabeling 
\begin{align}
 \label{relabelling}
 (z_{\phi_{1,1}}, z_{\phi_{1,2}}, \ldots, z_{\phi_{1,\chi_1}}) &\to (y_{\chi_1}, y_{\chi_1 -1}, \ldots, y_1) \\
 (z_{\phi_{2,1}}, z_{\phi_{2,2}}, \ldots, z_{\phi_{2,\chi_2}}) &\to (y_{\chi_1 + \chi_2}, y_{\chi_1 + \chi_2 -1}, \ldots, y_{\chi_1 + 1})\\ 
 &\vdots
 \end{align}
which we indicate shortly as $z_{\alpha,i} = y_{\phi^\ast(\alpha,i)}$; this defines implicitly, for each
$\phi$, the mapping $\phi^\ast: (\alpha,i) \to \{1,\ldots,\partN\}$.
We also label the final variables $w_j = y_{\chi_1 + \ldots + \chi_{j-1} + 1}$ 
for $1 \leq j \leq \nr{\chi}$. For instance, for the filling $\phi$ considered in \eqref{examplefilling}
\begin{equation}
\label{relabeling}
\raisebox{-15mm}{ \includegraphics[keepaspectratio=true,height=0.15\columnwidth]{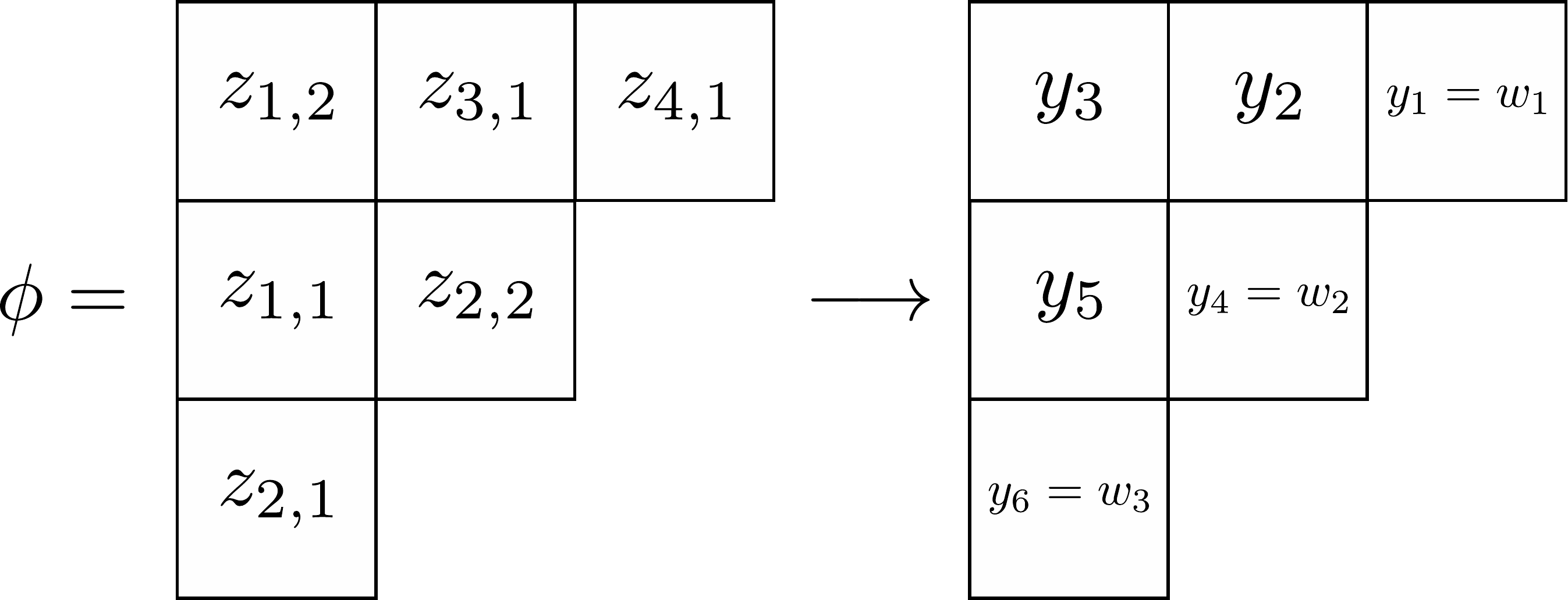}}\;.
\end{equation}
In this way, we arrive at 
\begin{multline}
 \label{step11}
 \mathcal{Z}_{\xi}(T) = \prod_{\alpha=1}^{\nc{\xi}} \frac{1}{\xi^\alpha!}
 \sum_{\stackrel{\chi \prec \xi}{\chi = 1^{a_1} 2^{a_2} \ldots}} \frac{1}{a_1! a_2!\cdots} 
 \sum_{\phi \in \Phi(\chi)} \int \cdots \int 
 \prod_{p=1}^{\nr{\chi}} {\frac{dw_p}{2 \pi i}}\\
 \times \Res_{\chi} \left[\left(\prod_{\alpha<\beta} \prod_{i,j} \frac{y_{\phi^\ast(\alpha,i)} - y_{\phi^\ast(\beta,j)}}{y_{\phi^\ast(\alpha,i)} - y_{\phi^\ast(\beta,j)} - 1}
 \right) \left(  \prod_{\alpha=1}^{\nc{\xi}} \prod_{i \neq j}^{\xi^\alpha} (y_{\phi^\ast(\alpha,i)} - y_{\phi^\ast(\alpha,j)}) 
 \right) F(\{y_{\phi^\ast}\})\right] \;.
\end{multline}
Notice that in this expression we replaced $\Res_\phi$ with $\Res_\chi$, since when written in the variables $y$'s,
the sequence of residues to be computed does not depend on the specific filling $\phi$ but only on the diagram $\chi$.
Now, we observe that if the constraint $\chi \prec \xi$
is removed and the sum is extended over arbitrary filling $\phi$ of the variables $y$'s in $\chi$, 
the result does not change.
This because the added terms will have a vanishing residue. 
For instance, if in \eqref{relabeling} one exchanges $z_{2,1}$ and $z_{1,1}$
there is no associated pole in the first factor. 
In this case the sum over all possible fillings $\phi$ without constraints, amounts
to a sum over all permutations of the variables $y$. Hence, the above expression can be replaced by a 
symmetrization over the variables $y$, 
and we arrive at 
\begin{equation}
 \label{step2}
\mathcal{Z}_{\xi}(T) = \prod_{\alpha=1}^{\nc{\xi}} \frac{1}{\xi^\alpha!}\sum_{\chi = 1^{a_1} 2^{a_2} \ldots} \frac{n!}{a_1! a_2!\cdots} 
 \int \cdots \int 
 \prod_{p=1}^{\nr{\chi}} {\frac{dw_p}{2 \pi i}}\\
 \times \Res_{\chi} \left[\sym_y W_0(\{y\})\right] \;.
\end{equation}
with $W_0(\{y\})$ defined as  
\begin{multline}
\label{W0def}
W_0 (\{y\}) \equiv 
\left(\prod_{\alpha<\beta} \prod_{i,j}\frac{y_{\phi^\ast(\alpha,i)} - y_{\phi^\ast(\beta,j)}}
 {y_{\phi^\ast(\alpha,i)} - y_{\phi^\ast(\beta,j)} - 1}
 \right) 
 \left( \prod_{\alpha=1}^{\nc{\xi}} 
  \prod_{i \neq j}^{\xi^\alpha} (y_{\phi^\ast(\alpha,i)} - y_{\phi^\ast(\alpha,j)}) 
 \right) F(\{y_{\phi^\ast}\})= \\
 =  \left(\prod_{k \neq l}\frac{y_k - y_l}{y_k - y_l - 1} \right) 
 \left(\prod_{k > l}\frac{y_k - y_l - 1}{y_k - y_l} \right) 
   \left( \prod_{\alpha=1}^{\nc{\xi}} 
  \prod_{i > j}^{\xi^\alpha} (y_{\phi^\ast_{\xi}(\alpha,i)} - y_{\phi^\ast_{\xi}(\alpha,j)})
  (y_{\phi^\ast_{\xi}(\alpha,j)} - y_{\phi^\ast_{\xi}(\alpha,i)}-1) 
 \right) F(\{y_{\phi^\ast_{\xi}}\})\;.
\end{multline}
Here, in the first line, because of the symmetrization
the filling $\phi$ of $\xi$ is arbitrary. In the second line of (\ref{W0def}),
and everywhere below, we choose the ``canonical'' one $\phi \equiv \phi_\diagr$ satisfying $\phi^\ast_\diagr(\alpha, j+1)-\phi^\ast_\diagr(\alpha, j)=1$, and increasing with $\alpha$ 
(see \eqref{phixiex} below for an explicit example),  
in agreement with the ordering in Fig.~\ref{fig:diagramSketch}  -- note: 
do not confuse the filling $\phi_\xi$ of the diagram $\xi$, used to define $W_0$ in \eqref{W0def},
with the the filling of $\chi$ in (\ref{relabeling}).
The only term singular when evaluating the residue
in this expression is the first one, which is already symmetric over $y$'s, and can be expressed
as a determinant of a $\nr{\chi}\times\nr{\chi}$ matrix \cite{borodin2014macdonald}
\begin{equation}
\Res_{\chi} \left[\prod_{k\neq l}\frac{y_{k} - y_{l}}{y_{k} - y_{l} - 1} \right] =
\det \left(\frac{1}{w_i + \chi_i - w_j}\right)_{i,j=1}^{\nr{\chi}}\;.
\end{equation}
The remaining term is not singular upon evaluation of these residues 
and we just need to insert the variables obtaining our main result:

\vspace{0.3cm}
\noindent \textbf{Proposition: ---}

\begin{equation}
 \label{finalborodin}
 \mathcal{Z}_{\xi}(T) = \prod_{\alpha=1}^{\nc{\xi}} \frac{1}{\xi^\alpha!}\sum_{\chi = 1^{a_1} 2^{a_2} \ldots} \frac{n!}{a_1! a_2!\cdots} 
 \int \cdots \int 
 \prod_{p=1}^{\nr{\chi}} {\frac{dw_p}{2 \pi i}}\\
 \times \det \left(\frac{1}{w_i + \chi_i - w_j}\right)_{i,j=1}^{\nr{\chi}} 
 \Sub_{\chi} \left[\sym_y W_\xi(\{y\})\right] \;.
\end{equation}
\textit{Here the operator $\Sub_{\chi}$ requires replacing  $y_i=y_{i-1}+1$ for 
for $i \nin \{1, 1 + \chi_1, 1+\chi_1 + \chi_2,\ldots \, ,1+\chi_1+\ldots+\chi_{n_R(\chi)-1}\}$ 
and considering the result as a function of $w_j = y_{1 + 
\sum_{l=1}^{j-1} \chi_l}$,  with
$j = 1,\ldots,n_R(\chi)$.
The function $W_\xi(\{y\})$ contains the non-symmetric part of $W_0(\{y\})$ 
and is the only part depending on the choice of the diagram $\xi$:
\begin{equation}
\label{bigWdef}
W_\xi (\{y\}) \equiv 
 \left(\prod_{k > l}\frac{y_k - y_l - 1}{y_k - y_l} \right) 
   \left( \prod_{\alpha=1}^{\nc{\xi}} 
  \prod_{i > j}^{\xi^\alpha} (y_{\phi^\ast_\xi(\alpha,i)} - y_{\phi^\ast_\xi(\alpha,j)})
  (y_{\phi^\ast_\xi(\alpha,j)} - y_{\phi^\ast_\xi(\alpha,i)}-1) 
 \right) F(\{y_{\phi^\ast_\xi}\})\;.
\end{equation}
with $\phi_\xi$ the ``canonical'' filling associated to the diagram $\xi$.}

\vspace{0.3cm}
\noindent Remarkably, all the other terms remain equal to the bosonic case. 
Finally, for the two-rows diagram $\xi = (\partN-\partM, \partM)$, Eq.~\eqref{QpExp} and \eqref{borodinG} are recovered
by the change of variables $w_j \to i k_j/\cbar - (m_j-1)/2$, $j=1,\ldots,\partNS$, and in general 
$y_A \to i \mu_A/\cbar$ for $A=1,\ldots,n$,  where each row of $\chi$ is identified with a string configuration, so that
$m_j = \chi_j$ and $\partNS = n_R(\chi)$. The substitutions in \eqref{rescfrombor} 
are finally applied.
For instance, for $\xi = (42)$, we have the filling
\begin{equation}
\label{phixiex}
 \phi_\xi = \raisebox{-10mm}{ \includegraphics[keepaspectratio=true,height=0.12\columnwidth]{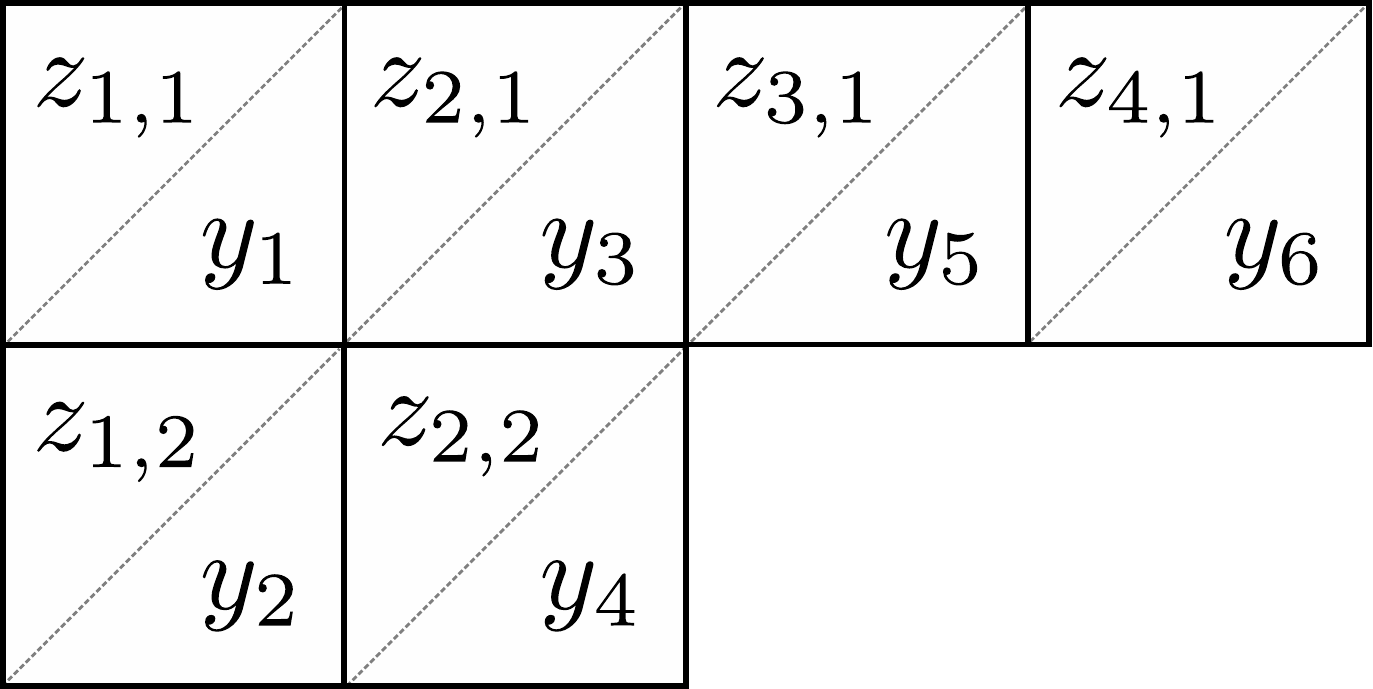}} 
\end{equation}
and we obtain 
\begin{equation}
  \prod_{\alpha, i > j} (y_{\phi^\ast_\xi(\alpha,i)} - y_{\phi^\ast_\xi(\alpha,j)})
  (y_{\phi^\ast_\xi(\alpha,j)} - y_{\phi^\ast_\xi(\alpha,i)}-1) = [(y_2 -y_1)(y_1 -y_2-1)][(y_4-y_3)(y_3 -y_4-1)]= \frac{h(\mu_{1,2})h(\mu_{3,4})}{\cbar^4} \;.
\end{equation}
To avoid confusion, we remark that Eq.~\eqref{phixiex} is used to define $W_\xi$ in \eqref{bigWdef}. Only after the symmetrization over $y$
in \eqref{finalborodin} has been performed, one must carry on the replacement 
$\Sub_\chi$ corresponding to each diagram $\chi$, as indicated above, which is
equivalent to injecting the string eigenstates. 

In general, the determinant in \eqref{finalborodin} produces the term $\Phi(\kv, \mv)$ in \eqref{QpExp}, while the first and second factor in \eqref{bigWdef}
give respectively the denominator and the numerator in \eqref{borodinG}.

\section{GGE partition function} 

We derive a Fredholm determinant form for the generalized generating function, defined as
\begin{equation}
 \label{genfunbeta}
 \genfun_{\bv}(u, t) = \sum_{\partN \geq 0} \frac{(-u)^{\partN}}{\partN!} \ZZ_\partN^{\bv} (t) \;.
\end{equation}
where, in this work, we introduced the GGE partition function as the following sum
restricted over {\it bosonic states} (i.e. fully symmetric)
\begin{equation}
\label{GGEpartfun}
\ZZ_{\partN}^{\bv}(t) 
= \sum_{\mu  \text{\tiny, bosonic}} \frac{|\psi_\mu(0)|^2}{||\mu||^2} e^{- t A_2 + \sum_{p \geq 1} \beta_p A_p} 
=  \sum_{\partNS=1}^\partN \frac{\partN! \cbar^\partN}{\partNS! (2 \pi \cbar)^{\partNS}} \sum_{(m_1,..m_{\partNS})_{\partN}}  
\prod_{j=1}^{\partNS} \int_{-\infty}^{+\infty} \frac{dk_j e^{-t A_2 + \sum_{p \geq 1} \beta_p A_p} } {m_j} 
\Phi(\kv, \mv) \;.
\end{equation}
the charges being defined as $A_p = \sum_j \mu_j^p$, and the
second equality uses the decomposition over string states valid in the 
large $L$ limit. It is thus a 
generalization of the bosonic partition sum of the fixed endpoint directed polymer 
studied in \cite{calabrese2010free,dotsenko}. Here it 
appears naturally in the non-symmetric problem, and its calculation
generalizes the one of Ref. \cite{calabrese2010free}. Note that in (\ref{genfunbeta}) we have
defined $\ZZ_\partN^{\bv} (t)=1$, which we find to be a consistent analytical continuation of the GGE
partition sum to $n=0$.  

After the insertion of the string ansatz, the charges take the form $A_p = \sum_{j=1}^{\partNS} \tilde{A}_p(k_j,m_j)$ 
with 
\begin{align}
\label{chargesstring}
\tilde{A}_1(k,m) &= k m \;,\\
\tilde{A}_2(k,m) &= m k^2 + \frac{1}{12} (m - m^3)\cbar^2 \;,\\
\tilde{A}_3(k,m) &= \frac{ k m }{4}\left(4 k^2 + (1-m^2)\cbar^2\right) \;.
\\
\tilde{A}_4(k,m)  & = m k^4 + \frac{1}{2} \bar c^2 k^2 m
   \left(1-m^2\right) + 
 \frac{1}{240} \bar c^4 m \left(3 m^4-10 m^2+7\right) 
\end{align}

Exchanging in \eqref{genfunbeta} the sum over $\partNS$ and the one over $\partN$, we can rewrite it as 
\begin{equation}
 \genfun_{\bv}(u, t) = \sum_{\partNS = 0}^\infty \frac{1}{\partNS!} \ZZ^{(\partNS)}_{\bv}(u, t)
\end{equation}
where the partition function with a fixed number of strings has been introduced
\begin{equation}
 \label{genZstrings}
\ZZ^{(\partNS)}_{\bv}(u, t) = \sum_{m_1\ldots m_{\partNS}=1}^\infty   \frac{(-u \cbar)^{\sum_j m_j}}{(2 \pi \cbar)^{\partNS}} 
\int_{\mathbb{R}^{\partNS}}\left[\prod_{j=1}^{\partNS} \frac{dk_j e^{-t \tilde{A}_2(k_j,m_j) + \sum_{p \geq 1} \beta_p \tilde{A}_p(k_j,m_j) }} {m_j} 
\right]\Phi(\kv, \mv) \;.
 \end{equation}
To proceed further we rewrite this expression introducing the determinant of a $\partNS\times\partNS$ matrix
\begin{equation}
\det  \left(\frac{1}{2 i k_i - 2 i k_j  + \cbar (m_i + m_j) }\right) \prod_{j=1}^{\partNS} (2 \cbar m_j)  =
\Phi(\kv, \mv)
\end{equation}
which follows by the Cauchy-determinant formula \cite{schechter1959inversion}, and leads to
\begin{equation}
 \label{genDet}
\ZZ^{(\partNS)}_{\bv} (u, t) = 2^{\partNS} \sum_{m_1\ldots m_{\partNS}=1}^\infty  
\int_{\mathbb{R}^{\partNS}} \left[\prod_{j=1}^{\partNS} \frac{dk_j (-u \cbar)^{m_j} e^{-t \tilde{A}_2(k_j,m_j) +  \sum_{p \geq 1} \beta_p \tilde{A}_p(k_j,m_j)}}{2\pi} \right]
\det  \left(\frac{1}{2 i k_i - 2 i k_j  + \cbar (m_i + m_j) }\right) 	\;.
 \end{equation}
We now employ the Laplace expansion for the determinant
\begin{multline}
\label{detlaplace}
\det  \left(\frac{1}{2 i k_i - 2 i k_j  + \cbar (m_i + m_j) }\right) = \sum_{P} (-1)^{\sigma_P} \prod_{j=1}^{\partNS} 
\frac{1}{2 i k_j - 2 i k_{P_j}  + \cbar (m_j + m_{P_j}) } = \\=
\sum_{P} (-1)^{\sigma_P} \int_{v_j>0} \prod_j dv_j e^{ -v_j (2 i k_j - 2 i k_{P_j}  + \cbar (m_j + m_{P_j}))} =
\sum_{P} (-1)^{\sigma_P} \int_{v_j>0} \prod_j dv_j e^{ - 2 i k_j (v_j - v_{P_j}) - \cbar m_j (v_j + v_{P_j})} 
\end{multline}
where in the last equality we exchanged $P$ with $P^{-1}$ taking into account that they have the same parity.
Replacing \eqref{detlaplace} inside \eqref{genDet} we obtain 
\begin{equation}
 \label{genDetFull}
\ZZ^{(\partNS)}_{\bv}(u, t) = \prod_{j=1}^{\partNS} \int_{v_j > 0} \det \kk^{\bv}_u (v_i, v_j).
 \end{equation}
with the Kernel
\begin{equation}
 \label{kernelM}
 \kk^{\bv}_u (v_1, v_2) = \int_{-\infty}^{\infty} \frac{dk}{2 \pi} 
 \sum_{m=1}^\infty 2(-\cbar u)^m e^{-t \tilde{A}_2(k,m) + \sum_{p \geq 1} \beta_p \tilde{A}_p(k,m) }  
 e^{-2 i k (v_1-v_2) - \cbar m (v_1+v_2)} \;.
\end{equation}
From this it follows that
\begin{equation}
 \label{FredholmGen}
\genfun_{\bv}(u, t) = \Det ( 1 + \Proj_0 \kk^{\bv}_u \Proj_0)
 \end{equation}
 in terms of a Fredholm determinant, where $\Proj_s$ is the projector on $[s, \infty)$.

\section{Moments from generating function}

As discussed in the text we expand the expressions for $\QQ_{\partN,m}$ for $m=1,2,3$ 
in the conserved charges of the model $A_p$. We display here their full expression
for $m=1,2$ and only the leading order in $1/n$ for $m=3$:
\begin{subequations}
\label{expr1}
\begin{align}
\label{m1}
&\QQ_{\partN,1} = \frac{1}{n(n-1)} \left( n A_2 - A_1^2 + \frac{n^2 (n^2-1)}{12} \cbar^2 \right) = \frac{A_1^2}{n} + (A_1^2-A_2) + O(n)\;,  \\
\label{m2}
&\QQ_{\partN,2} = 
\frac{ (\partN^2-1) \partN (5 \partN-6) \bar c^4}{720} -\frac{(A_1^2-\partN A _2) \bar c^2}{6}   + \frac{4 (\partN-1)
A_3 A _1+(\partN^2-3\partN+3) A _2^2-(\partN-1) \partN A _4+A _1^4 -2 \partN A _2 A _1^2}{(\partN-3) (\partN-2) (\partN-1) \partN} = \\
& \specialcell{\hfill = \frac{1}{6 n} (4 A_1 A_3 - A_1^4 - 3 A_2^2) \;,
+ \frac{1}{36} ( 12 A_1^2 A_2 + 20 A_1 A_3 - 11 A_1^4 - 15 A_2^2 - 6 A_4) - \frac{A_1^2}{6} \bar c^2+ O(n)} \label{m22} \\
\label{m3}
&\begin{multlined}[b][.87\textwidth]
\Lambda_{n,3}= \frac{1}{120 n} \left[-10 \left(A_1^4-4 A_3 A_1+3 A_2^2\right) \bar c^2+A_1^6-20 A_3 A_1^3+15
   \left(A_2^2+2 A_4\right) A_1^2+  \right.\\ 
   \left. +\left(24 A_5-60 A_2
   A_3\right) A_1+10 \left(3 A_2^3-6 A_4 A_2+4
   A_3^2\right)\right] + O(n^0) \;.
\end{multlined}
\end{align}
\end{subequations}
Using the GGE partition function in \eqref{GGEpartfun}, we can rewrite
\begin{equation} \label{sub} 
 \Theta_{\partN,m}(t)= \QQ_{\partN,m}(\{\partial_p\}) [ \ZZ_\partN^{\bv} (t) ]
\end{equation}
where we replaced in $\QQ_{\partN,m}$ the charges $A_p \to \partial_{p}$ computed 
setting all $\beta$'s to zero afterwards. We immediately see that the first
expression (\ref{m1}) for $\QQ_{\partN,1}$ leads to equation \eqref{eq1} in the text,
valid for any $n$, using the equivalence $\partial_2 \equiv - \partial_t$
and replacing $A_1^2 \to \frac{n}{2 t}$ which comes from the STS (see below). 
Note also that in the limit $n \to 0$ we will be allowed to discard in the Taylor expansion as written in
(\ref{m1}-\ref{m3}), the terms formally of order 
$O(n^0)$ and higher (not written for $m=3$), since (i) we have checked that they
all contain derivatives acting on $\ZZ_\partN$ 
and (ii) $\mathcal{Z}_\partN^{\bv}(t) \to 1$ when $n\to 0$ (see above).

The above expressions can be further simplified employing the STS symmetry, manifested here as
the invariance of \eqref{GGEpartfun}
under the shift of all rapidities $\mu_j \to \mu_j + k$ for an arbitrary constant $k$. 
Such symmetry was used in e.g. Appendix of Ref. \cite{flat} in presence
of charges $A_1, A_2$ only. Here 
we extend the analysis to the enlarged context of the GGE partition
sum, which contains all charges $A_p$.
Since under the shift
($A_0\equiv n$)
\begin{equation}
 A_p \to A_p + \sum_{q=1}^{p} \binom{p}{q} A_{p-q} k^q 
\end{equation}
at the order $O(k)$ in \eqref{GGEpartfun}, we obtain the equality, valid for arbitrary $\bv$
\begin{equation}
\left[-2 t \partial_1 + n  \beta_1 + \sum_{p=2}^\infty p \beta_p \partial_{p-1} \right] \ZZ_{\partN}^{\bv} (t) = 0\;.
\end{equation}
By expansion order by order in $\bv$, one obtains an infinite list of equalities involving conserved charges, that can 
be generated easily by
\begin{equation}
\langle  \left[-2 t A_1 + \sum_{p=1}^\infty p A_{p-1} \partial_{A_p} \right] q(\{A\}) \rangle  ~ = 0\;.
\end{equation}
where $q(\{A\})$ is an arbitrary polynomial in the $A_p$. Here,
these identities are understood as inserted in the integral (\ref{QpExp})
over string momenta, which is denoted here as $\langle \ldots \rangle$.
Equivalently, the charges are afterwards replaced by derivatives 
applied to $\ZZ_\partN^{\bv} (t)$ as in \eqref{sub}. Varying the choice of $q(\{A\})$, we obtain for instance
\begin{equation}
\langle A_1^{2k+1} \rangle  = 0, \qquad \langle A_1^{2k} \rangle   = \frac{n^k (2k-1)!! }{2^k t^k},\quad \langle A_1 A_3 \rangle   = \frac{3 \langle A_2 \rangle }{2t}\;. \label{ids} 
\end{equation}
where the last equality goes beyond the standard use of STS (in e.g. Ref. \cite{flat}). 

Thanks to these identities, the limit $\partN \to 0$ can be taken in \eqref{expr1}. For
$m=1,2$ only $A_2$ survives, that, being the energy, can be replaced according to $\partial_2 \equiv - \partial_t$.
Then using (\ref{ids}), Eqs. \eqref{m1} and \eqref{m22} lead respectively to \eqref{eq1} 
and \eqref{p2eq} in the text, using that 
\begin{equation}
\frac{1}{n} \partial_2^p \ZZ_\partN^{\bv} (t)|_{\bv=0} 
=  \frac{1}{n}  (-\partial_t)^p  {\cal Z}_\partN (t) \xrightarrow{n \to 0} (-\partial_t)^p \overline{ \ln Z_\eta(0;0|t) }\;.
\end{equation}
For $m\geq 3$, higher conserved charges are involved and the moments are not simply related to 
the free-energy distribution. 
For instance for $m=3$ after using STS we find that we can replace:
\begin{equation}\
\QQ_{n,3} \to \frac{1}{\partN} \left[ \frac{\cbar^2 A_2}{2 t} - \frac{\cbar^2 A_2^2}{4}  - \frac{3 A_2^2}{4 t}  + \frac{A_2^3}{4}  + \frac{A_3^2}{3}  + 
\frac{ A_4 }{2 t}- \frac{A_2 A_4 }{2}  \right]  \label{m3new} 
\end{equation}
with no further simplification. Using (\ref{sub}) the final result for $\overline{ p_\eta(t)^3}$ is thus a linear combination of
\begin{equation}
\label{rhodef}
 \rho_{i_1,\ldots,i_k} \equiv \left.\lim_{\partN \to 0} \partial_{i_1}\ldots \partial_{i_k} \frac{\mathcal{Z}_\partN^{\bv}(t) - 1}{\partN}\right|_{\bv = 0} \;.
\end{equation}
where $\partial_{i_1} \equiv \partial_{\beta_{i_1}}$. The 
$\rho_{i_1,\ldots,i_k}$ can then be expressed as derivatives acting on 
the Fredholm determinant expression\eqref{FredholmGen}. To see this, we write it as 
a Mellin transform
\begin{equation}
 \label{Zfourier}
 \mathcal{Z}_{\partN}^{\bv} (t) = \int_0^\infty dy \tilde{\mathcal{Z}}_{\bv} (y, t) y^{\partN}\;,
\end{equation}
which can be inserted in  \eqref{genfunbeta} to give
\begin{equation}
 \genfun_{\bv}(u,t) = \int_0^\infty dy \tilde{\mathcal{Z}}_{\bv} (y, t) e^{-uy}\;.
\end{equation}
with $\genfun_{\bv}(0,t)=1$. 
This shows that $\tilde{\mathcal{Z}}_{\bv} (y, t)$ can be equivalently defined as the inverse Laplace transform of $\genfun_{\bv}(u,t)$.
Inserting \eqref{Zfourier} in \eqref{rhodef} and taking the limit $\partN \to 0$, we obtain for $k>0$
\bea
 \rho_{i_1,\ldots,i_k} &= & \partial_{i_1}\ldots \partial_{i_k}  \int_0^\infty dy \tilde{\mathcal{Z}}_{\bv} (y, t) \ln y
  = 
  \partial_{i_1}\ldots \partial_{i_k}  \int_0^\infty \frac{du}{u} \left(e^{-u}-J_{\bv}(u, t)\right) = 
   -\int_0^\infty \frac{du}{u} \partial_{i_1}\ldots \partial_{i_k} J_{\bv}(u, t)\; \nonumber \\
&=&  -\int_0^\infty \frac{du}{u} \partial_{i_1}\ldots \partial_{i_k} \Det ( 1 + \Proj_0 \kk^{\bv}_u \Proj_0) . 
\eea
Derivatives of a FD are easily evaluated, for instance for $k=2$ one obtains
\begin{multline}
\label{FredholmDer}
\partial_a \partial_b J_{\bv}(u, t) = \left[\Tr[ (1 + \kk^{\bv}_u )^{-1} \partial_a \partial_b \kk^{\bv}_u  ]+\Tr[ (1 + \kk^{\bv}_u)^{-1} \partial_a  \kk^{\bv}_u  ]\Tr[ (1 + \kk^{\bv}_u)^{-1} \partial_b  \kk^{\bv}_u  ] + \right. \\ 
\left.- \Tr[ (1 + \kk^{\bv}_u)^{-1} \partial_a \kk^{\bv}_u (1 + \kk^{\bv}_u)^{-1} \partial_b \kk^{\bv}_u  ]\right] \Det(1 + \Proj_0 \kk^{\bv}_u \Proj_0) \;.
\end{multline}
The derivatives with respect to the Lagrange multiplier $\bv$, once applied to the kernel $\kk^{\bv}_u$
can be converted, using \eqref{kernelM} and \eqref{chargesstring}, into a combination of derivatives with respect to $v_1, v_2$, e.g.
\begin{align}
 \label{chargederivative}
 \partial_1 \kk^{\bv}_u(v_1, v_2) &= \frac{\partial_{v_1 - v_2} \partial_{v_1 + v_2} \kk^{\bv}_u(v_1, v_2) }{2 i \cbar} = -\frac{i (\partial_{v_1}^2 - \partial_{v_2}^2) \kk^{\bv}_u(v_1, v_2) }{8 \cbar} \\
 \partial_3 \kk^{\bv}_u(v_1, v_2) &= \frac{i(\partial_{v_1}^4 - \partial_{v_2}^4) \kk^{\bv}_u(v_1, v_2)}{64 \cbar} + \frac{\cbar^2 \partial_1 \kk^{\bv}_u(v_1, v_2)}{4} \;.
\end{align}
and more generally, from \eqref{kernelM}
\begin{equation}
\partial_j \kk^{\bv}_u(v_1, v_2) = \tilde A_p(k = \frac{i}{2} \partial_{v_1-v_2} , 
m = - \frac{1}{\bar c} \partial_{v_1+v_2}) \;.
\end{equation}
These formulas can be iterated in case of multiple derivatives, leading to explicit but lengthy expressions.
The limit $\bv \to 0$ can now be taken safely and we then have to deal only with the standard kernel
$\kk_u \equiv \kk^{\bv = 0}_u$, as given in Ref. \cite{calabrese2010free}. 
The calculation at arbitrary time by this method is possible but very demanding.

Let us now show how the asymptotics at large time simplifies.
We set $\lambda^3 = \cbar^2 t/4$ and introduce $s$ by $ e^{-\lambda s}=(u \cbar) e^{-\frac{\lambda^3}{3}}$.
Moreover, we rescale $k \to k \cbar/\lambda$ and $v_{1,2} \to \lambda v_{1,2}/\cbar$, 
so that $\kk_u(v_1, v_2) \to \lambda\kk_s (v_1, v_2)/\cbar $, because of the integration measure, and we obtain finally
the $\lambda \to \infty$ limit
\begin{equation}
\label{kernelLargeTime}
\kk_s^{\infty}(v_1, v_2) =  -\int_{-\infty}^{\infty} \frac{dk}{2 \pi} \int_{0}^\infty dy \Ai(y +  k^2+s+v_1+v_2) 
 e^{-i k (v_1-v_2)} = - 2^{1/3} K_{\Ai}(2^{1/3} (v_1+ s/2), 2^{1/3} (v_2+ s/2)) \;.
\end{equation}
with the Airy Kernel defined as
\begin{equation}
 K_{\Ai}(x,y) = \int_0^\infty dy \Ai(v_1 + y) \Ai(v_2 + y) = \frac{\Ai(v_1) \Ai'(v_2) - \Ai'(v_1) \Ai(v_2)}{v_1 - v_2} \;.
\end{equation}
In this regime, the Fredholm determinant can be computed  efficiently numerically
following \cite{bornemann2009numerical}.

Let us show how to obtain the leading large time behavior on the example of $m=3$. 
First one sees that one can decompose the charges (\ref{chargesstring}) as:
\bea
&& \tilde A_2(k,m) = \hat A_2(k,m) + \frac{\bar c^2}{12} m\;,  \\
&& \tilde A_3(k,m)  = \hat A_3(k,m)   + \frac{\bar c^2}{4} \tilde A_1(k,m)\;, \\
&& \tilde A_4(k,m)  = \hat A_4(k,m)   + \frac{7 \bar c^4}{240} m +  \frac{\bar c^2}{2} \hat A_2(k,m)
\eea
and $\tilde A_1(k,m) = \hat A_1(k,m)$, where $\hat A_p(k,m)$ is the homogeneous part, i.e. $\hat A_p( \frac{k\cbar }{\lambda} ,m)= \lambda^{-(p+1)}
\hat A_p(k \cbar, \lambda m)$ under the rescaling $k \to k \cbar/\lambda$.
Thus in (\ref{m3new}) one can formally decompose the charges as:
\bea
\label{homocharg}
A_2 = \hat A_2 + \frac{\bar c^2 n}{12} \quad , \quad A_3 = \hat A_3 + \frac{\bar c^2 A_1}{4} \quad , \quad 
A_4 = \hat A_4 + \frac{7 \bar c^4 n}{240} +  \frac{\bar c^2 \hat A_2 }{2} 
\eea 
and the above analysis of the FD at large time shows that the hat charges scale as
$\hat A_p \sim t^{-(p+1)/3}$, which we denote by "normal" scaling, 
and ${\cal Z}_n/n$ scales as $t^{1/3}$. 
For example $A_2 \mathcal{Z}_n/n \xrightarrow{n \to 0}  - \partial_t \overline{\ln Z_\eta(0;0|t)}
= \bar c^2/12 + O(t^{-2/3})$ has ``anomalous'' scaling, while $\hat A_2 \mathcal{Z}_n/n \xrightarrow{n \to 0} O(t^{-2/3})$ 
has ``normal'' scaling with time at large time. 

Putting all together, in the limit $n \to 0$ we find:
\bea
\overline{p_\eta(t)^3}  \simeq \frac{\cbar^4}{15 t} -  \frac{2 \overline{\chi_2} \cbar^{8/3}}{9 t^{5/3}}
\eea 
where the sub-leading term is obtained using \eqref{homocharg} in \eqref{m3new}. Then the term scaling as $t^{-5/3}$ gives at large times 
\begin{equation}
 \lim_{n\to 0} \left(\frac{3 \hat A_2 }{4 t} - \frac{\hat A_2^2 }{2} + \frac{ \hat A_1 \hat A_3}{6}\right) \frac{\cbar^2 \mathcal{Z}_n}{n}
 = -\frac{\cbar^4}{12 t} - \cbar^2 \left(\frac{\partial_t}{t} + \frac{\partial_t^2}{2}\right) h(t) = -  \frac{2 \overline{\chi_2} \cbar^{8/3}}{9 t^{5/3}}\;.
 \end{equation}

\section{conjecture for $\overline{\ln p_\eta(t)}$ }

In \cite{Doumerc}
relations between some discrete DP models and the eigenvalues 
of the GUE (and LUE) random matrix ensembles are discussed. 
For instance, calling $h_k$ 
the maximum energy of an ensemble of $k$ non-crossing paths of $\simeq N$ steps with
nearest neighbor endpoints 
in a semi-discrete DP model, and $\lambda_1>\lambda_2$ the two
largest eigenvalues of the GUE(N), it is stated that:
\bea
&& h_1 =_d \lambda_1 \\
&& h_2 - h_1 =_d \lambda_2 
\eea 
where $=_d$ denotes equality in law (where, for our purpose, we consider $N$ large). Under appropriate rescaling
\cite{borodin2014macdonald}, universality then suggests that at 
large time for the continuum DP model:
\bea
&& \ln Z_\eta \simeq -\frac{\cbar^2 t}{12} + \overline{\chi_2}(\cbar^2 t)^{1/3} \\
&& \ln \frac{Z^{(2)}_\eta(\epsilon)}{(2\epsilon)^2}  - \ln Z_\eta
\simeq -\frac{\cbar^2 t}{12} + \overline{\chi'_2}(\cbar^2 t)^{1/3}
\eea 
where the limit of small $\epsilon$ is implicit, and
we denote $Z_\eta\equiv Z_\eta(0;0|t)$. Here
$\chi_2 > \chi'_2$ are the two
largest eigenvalues of the GUE, scaled so that
the largest one obeys the Tracy Widom cumulative distribution $F_2$. These equations 
imply:
\bea
&&\overline{\ln p_\eta(t)} = \overline{\ln \frac{Z^{(2)}_\eta(\epsilon)}{(2\epsilon)^2}} - 2 ~ \overline{ \ln Z_\eta} \simeq
- ( \overline{\chi_2} - \overline{\chi'_2} ) (\cbar^2 t)^{1/3} \approx - 1.9043 (\cbar^2 t)^{1/3}
\eea 
with $\overline{\chi_2}= - 1.7710868$ and $\overline{\chi'_2}=- 3.6754$
(value given in \cite{TWNote}). We have obtained preliminary numerical
support for this behavior \cite{delucaledou}. 
This means that in typical disorder
configurations $p_\eta(t)$ decreases (sub)-exponentially fast with time, while its moments
are dominated by rare, but not so rare, disorder configurations. This is similar
to the behavior of the probability $q_\eta(t)$ that a single DP does not cross a hard wall
at $x=0$ studied in \cite{Gueudre}, where it was found that
$\overline{\ln q_\eta(t)} = - ( \overline{\chi_2} - \overline{\chi_4} ) (\cbar^2 t)^{1/3}
\approx - 1.49134 (\cbar^2 t)^{1/3}$
where $\chi_4$ is distributed according to the GSE TW distribution
with $\overline{\chi_4}=- 3.2624279$. Note that avoiding a second polymer
(which itself wanders and competes for the same favorable regions of the random potential) 
is more restrictive in phase space for the DP than a hard wall, consistent
with $\overline{\ln p_\eta(t)}  < \overline{\ln q_\eta(t)}$ as we find.

%

\end{document}